\documentclass[acmsmall]{acmart}

\setcopyright{none}

\usepackage{booktabs}   
\usepackage{subcaption} 

\usepackage{lipsum}
\usepackage{hyperref}
\usepackage{textgreek}
\usepackage{amsmath}
\usepackage{amssymb}
\usepackage{bbm}
\usepackage[greek,english]{babel}
\usepackage[utf8]{inputenc}
\usepackage{url}
\usepackage{proof}
\usepackage{fancyvrb}

\DefineVerbatimEnvironment{verbatim}{Verbatim}{xleftmargin=.2in}

\DeclareUnicodeCharacter{25CF}{\ensuremath{\blacksquare}}
\DeclareUnicodeCharacter{2200}{\ensuremath{\forall}}
\DeclareUnicodeCharacter{2605}{\ensuremath{\star}}
\DeclareUnicodeCharacter{2794}{\ensuremath{\rightarrow}}
\DeclareUnicodeCharacter{2243}{\ensuremath{\simeq}}
\DeclareUnicodeCharacter{03A0}{\ensuremath{\mathrm{\Pi}}}
\DeclareUnicodeCharacter{03BB}{\textlambda}
\DeclareUnicodeCharacter{923}{\textLambda}
\DeclareUnicodeCharacter{25C2}{\ensuremath{\blacktriangleleft}}
\DeclareUnicodeCharacter{27BE}{\ensuremath{\Rightarrow}}
\DeclareUnicodeCharacter{2218}{\ensuremath{\circ}}
\DeclareUnicodeCharacter{2081}{$_1$}
\DeclareUnicodeCharacter{2082}{$_2$}
\DeclareUnicodeCharacter{7522}{$_i$}

\newcommand{\refsec}[1]{Section \ref{sec:#1}}
\newcommand{\labsec}[1]{\label{sec:#1}}
\newcommand{\reffig}[1]{Figure \ref{fig:#1}}
\newcommand{\labfig}[1]{\label{fig:#1}}

\newcommand{\abs}[4]{{#1}\, #2\! : \! #3.\, #4}
\newcommand{\lam}[2]{\lambda\, #1.\, #2}

\newcommand{\txt}[1]{\ensuremath{\texttt{#1}}}
\newcommand{\arr}[0]{\ensuremath{\rightarrow}}
\newcommand{\nat}[0]{\ensuremath{\mathbb{N}}}

\begin{document}

\title{Generic Zero-Cost Reuse for Dependent Types}


\author{Larry Diehl}
\affiliation{
  \institution{University of Iowa}            
  \country{USA}                    
}
\author{Denis Firsov}
\affiliation{
  \institution{University of Iowa}            
  \country{USA}                    
}
\author{Aaron Stump}
\affiliation{
  \institution{University of Iowa}            
  \country{USA}                    
}

\begin{abstract}
Dependently typed languages are well known for having a problem with
code reuse. Traditional non-indexed algebraic datatypes (e.g. lists)
appear alongside a plethora of indexed variations (e.g. vectors).
Functions are often rewritten for both non-indexed and indexed versions of
essentially the same datatype, which is a source of code duplication.

We work in a Curry-style dependent type theory, where the same untyped
term may be classified as both the non-indexed and indexed versions of
a datatype. Many solutions have been proposed for the problem of
dependently typed reuse, but we exploit Curry-style type theory in our
solution to not only reuse data and programs, but do so at zero-cost
(without a runtime penalty). Our work is an exercise in dependently
typed generic programming, and internalizes the process
of zero-cost reuse as the identity function in a
Curry-style theory.
\end{abstract}



\maketitle

\section{Introduction}

Dependently typed languages
(such as Agda~\cite{lang:agda}, Coq~\cite{lang:coq},
Idris~\cite{lang:idris}, or Lean~\cite{lang:lean}) can be used to
define ordinary algebraic datatypes, as well as indexed versions of
algebraic datatypes that enforce various correctness properties.
For example, we can index lists by natural numbers to
enforce that they have a particular length
(i.e. \texttt{$\txt{Vec}_A : \nat \arr \star $}).
Similarly, we can index lists by two elements to
enforce that they are ordered and have a lower and upper bound
(i.e. \texttt{$\txt{OList}_{A,R} : A \arr A \arr \star $}).
We can even combine these two forms of indexing to
enforce that lists have all of the aforementioned correctness properties
(i.e. \texttt{$\txt{OVec}_{A,R} : A \arr A \arr \nat \arr \star$}).

Which datatype a programmer uses depends upon how much correctness
they wish to enforce at the time a function is written, versus proving
correctness as a separate step sometime later (corresponding to
intrinsic and extrinsic correctness proofs of functions). Certain
types tend to be better suited to writing intrinsically correct
functions than others, e.g. it is natural to define a safe \txt{lookup}
function that takes a \txt{Vec} as an argument, and a correct \txt{sort}
function that returns an \txt{OList}.

However, once we have written a function using a suitable indexed
variant of a datatype, reusing the function to define a corresponding version
for the unindexed (or less indexed) datatype can be
painful. We may also wish to delay extrinsic verification by choosing
to pay the price later, i.e. when we reuse a function over unindexed (or less
indexed) datatypes to define a corresponding function over
more indexed datatypes. We refer to the former direction as
\textit{forgetful reuse}, and to the latter direction as
\textit{enriching reuse}.

One source of pain is \textit{manually writing} functions
over some datatypes by reusing functions over differently indexed
variants of the same underlying datatypes. Another source of pain is
that reusing functions involves linear-time conversions
between differently indexed types, resulting in a
runtime \textit{performance penalty} incurred by practicing the good software
engineering practice of code reuse. \footnote{
  Wherever we say linear-time conversions, we are assuming reasonable
  implementations of the conversions. Of course, the time complexity
  can be worse for poorly implemented conversions.
}
In this paper we address both of
these problems, for both the forgetful and enriching directions of
reuse, by:
\begin{enumerate}
\item Defining generic combinators to incrementally attack the problem
  of reuse for various types, where each combinator application
  results in simplified subgoals (similar to tactics).
\item Ensuring that the combinators are closed operations with respect to
  a type abstraction, which can be eliminated to obtain reused
  functions at zero-cost (i.e. no performance penalty).
\end{enumerate}
Our \textbf{primary contributions} are:
\begin{enumerate}
\item{\refsec{prog:fog}:} Generic combinator solutions to zero-cost
  \textit{forgetful program reuse}
  (combinator \verb;allArr2arr;, handling the type of non-dependent functions),
  and \textit{proof reuse} (combinator \verb;allPi2pi;,
  handling the type of dependent functions).
\item{\refsec{prog:enr}:} Generic combinator solutions to zero-cost
  \textit{enriching program reuse}
  (combinator \verb;arr2allArrP;, handling the type of non-dependent functions),
  and \textit{proof reuse} (combinator \verb;pi2allPiP;,
  handling the type of dependent functions).
\item{\refsec{data:fog}:} A generic combinator solution to zero-cost
  \textit{forgetful data reuse}
  (combinator \verb;ifix2fix;, handling the type of fixpoints for
  generically encoded datatypes).
\item{\refsec{data:enr} \& \refsec{rel:penr}:} Generic combinator solutions to zero-cost
  \textit{enriching data reuse}
  (combinators \verb;fix2ifix; and \verb;fix2ifixP;, handling the type of fixpoints for
  generically encoded datatypes).
\end{enumerate}

Our work has two notable limitations:
\begin{enumerate}
\item We can only perform zero-cost conversions when indexed types
  store their index data as \textit{erased} arguments
  (or more accurately, when the arguments of
  non-indexed and indexed types match after erasure). This may not
  always be desirable, and has consequences such as needing to
  recompute the length of a vector if it is needed dynamically.
\item Forgetful program (or proof) reuse is not possible when the domain of the
  indexed function is constrained, such as the vector head function, when using
  our \verb;allArr2arr; combinator. We discuss this further in
  \refsec{others:interop}, including how a small extension of our work
  (i.e. also constraining the domain of the non-indexed function)
  could partially address this problem.
\end{enumerate}

The remainder of our paper proceeds as follows:
\begin{itemize}
\item{\refsec{back}:} We review background material, covering the
  Curry-style type theory that our results are developed within, and
  providing intuition for why zero-cost conversions are motivated by
  Curry-style type theory.
\item{\refsec{prob}:} We explain the primary problems
  (linear-time reuse of programs, proofs, and data) we are solving
  through concrete examples, and provide manual solutions (zero-cost,
  or constant time, reuse), which our
  primary contribution combinators generalize via generic programming.
\item{\refsec{prog}:} We generically solve the problems of
  (both forgetful and enriching) zero-cost
  program and proof reuse (as combinators for the types of non-dependent
  and dependent functions).
\item{\refsec{data}:} We generically solve the problems of
  (both forgetful and enriching) zero-cost
  data reuse (as combinators for the type of fixpoints).
\item{\refsec{rel}:} We evaluate our work on a more complex example of
  reuse between unchecked and checked STLC terms, and also extend our
  \textit{functional} data reuse enrichment to the \textit{relational}
  setting, where there is a premise on the non-indexed data.
\item{\refsec{others}: We compare what we have done with related
  work. This includes comparing our results with the closely related
  work of dependently typed reuse via
  ornaments~\cite{ornaments:original} and
  dependent interoperability~\cite{dagand:interop}, the primary
  difference being that our work achieves \textit{zero-cost} reuse.}
\item{\refsec{future}: We go over extensions that we have already made
  to our work, not covered herein, as well as planned future work.}
\end{itemize}
All of our results have been formalized in
Cedille~\cite{stump17a,stump18,lang:cedille}, a dependently typed language
implementing the theory we work in
(covered in \refsec{back:cdle}).\footnote{
  \raggedright{The Cedille formalization accompanying this paper is available at:\\
  \url{https://github.com/larrytheliquid/generic-reuse}}
}

\section{Background}
\labsec{back}

\subsection{The Type Theory (CDLE)}
\labsec{back:cdle}

We briefly summarize the type theory, the Calculus of Lambda
Eliminations (CDLE), that the results of this paper depend on.  For
full details on CDLE, including semantics and soundness results,
please see the previous papers~\cite{stump17a,stump18,lang:cedille}.  The main
metatheoretic property proved in the previous work is logical
consistency: there are types which are not inhabited.  Cedille is an
implementation of CDLE, and all the code appearing in this paper is
Cedille code.

CDLE is an extrinsic (i.e. Curry-style)
type theory, whose terms are exactly those of the pure untyped lambda
calculus (with no additional constants or constructs).  The
type-assignment system for CDLE is not subject-directed, and thus
cannot be used directly as a typing algorithm.  Indeed, since CDLE
includes Curry-style System F as a subsystem, type assignment is
undecidable~\cite{Wells99}.  To obtain a usable type theory, Cedille
thus has a system of annotations for terms, where the annotations
contain sufficient information to type terms algorithmically.  But
true to the extrinsic nature of the theory, these annotations play no
computational role.  Indeed, they are erased both during compilation
and before formal reasoning about terms within the type theory, in
particular by definitional equality (see \reffig{cdle}).

\begin{figure}
\centering
\minipage[b]{0.72\textwidth}  
  \[
  \begin{array}{cc}
    \infer{\Gamma\vdash \abs{\Lambda}{x}{T'}{t} : \abs{\forall}{x}{T'}{T}}{\Gamma,x:T'\vdash t : T & \!\!x\not\in\textit{FV}(|t|)} & 
    \infer{\Gamma\vdash t\ -t' : [t'/x]T}{\Gamma\vdash t : \abs{\forall}{x}{T'}{T} & \Gamma\vdash t':T'} \\ \\

    \infer{\Gamma\vdash \beta : t \simeq t}{\Gamma\vdash t : T} &
    \infer{\Gamma\vdash \rho\ q\ -\ t : [t_2/x]T}{\Gamma\vdash q : t_1 \simeq t_2 & \Gamma \vdash t : [t_1/x]T} \\ \\

    \infer{\Gamma\vdash \phi\ q\ -\ t_1 \{t_2\} : T}{\Gamma\vdash q : t_1 \simeq t_2 & \Gamma \vdash t_1 : T} &
    \infer{\Gamma\vdash [t_1,t_2] : \abs{\iota}{x}{T}{T'}}{\Gamma\vdash t_1 : T & \Gamma\vdash t_2 : [t_1/x]T' & |t_1| = |t_2|} \\ \\

    \infer{\Gamma\vdash t.1 : T}{\Gamma\vdash t : \abs{\iota}{x}{T}{T'}} &
    \infer{\Gamma\vdash t.2 : [t.1/x]T'}{\Gamma\vdash t : \abs{\iota}{x}{T}{T'}} \\ \\

  \end{array}
  \]
\endminipage\hfill 
\minipage[b]{0.28\textwidth}
\[
  \begin{array}{lll}
    |\abs{\Lambda}{x}{T}{t}| & = & |t| \\
    |t\ -t'| & = & |t| \\
    |\beta| & = & \lam{x}{x} \\
    |\rho\ q\ - \ t| & = & |t| \\
    |\phi\ q\ - \ t_1 \{t_2\}| & = & |t_2| \\
    |[t_1,t_2]| & = & |t_1| \\
    |t.1| & = & |t| \\
    |t.2| & = & |t| 
  \end{array}
  \]
\endminipage
\caption{Introduction, elimination, and erasure rules for additional type constructs.}
\labfig{cdle}
\end{figure}

CDLE extends the (Curry-style) Calculus of Constructions (CC) with
implicit products, primitive heterogeneous equality,
and intersection types:
\begin{itemize}
\item \verb;∀ x : T. T';, the implicit product type of
  \citet{miquel01}.  This can be thought of as the type for
  functions which accept an erased input of type \verb;x : T;, and
  produce a result of type \verb;T';. There are term constructs
  \verb;Λ x. t; for introducing an implicit input \verb;x;, and
  \verb;t -t'; for instantiating such an input with \verb;t';. The
  implicit arguments exist just for purposes of typing so that they
  play no computational role and equational reasoning happens on terms
  from which the implicit arguments have been erased.

\item \verb;t₁ ≃ t₂;, a Curry-style heterogeneous equality type.  The terms
  \verb;t₁; and \verb;t₂; are required to be typed, but need not have
  the same type.  We introduce this with a constant \verb;β; which
  erases to \verb;λ x. x; (so our type-assignment system has no
  additional constants, as promised); \verb;β; proves \verb;t ≃ t; for
  any typeable term \verb;t;.  Combined with definitional equality,
  \verb;β; proves \verb;t₁ ≃ t₂; for any $\beta$-equal \verb;t₁; and
  \verb;t₂; whose free variables are all declared in the typing
  context.  We eliminate the equality type by rewriting, with a
  construct \verb;ρ q - t;.  Suppose \verb;q; proves \verb;t₁ ≃ t₂;
  and we synthesize a type \verb;T; for \verb;t;, where \verb;T; has
  several occurrences of terms definitionally equal to \verb;t₁;.
  Then the type synthesized for \verb;ρ q - t; is \verb;T; except with
  those occurrences replaced by \verb;t₂;.
  The construct \verb;φ q - t₁{t₂}; casts a term \verb;t₂; (of any type)
  to type \verb;T;, provided that \verb;t₁; has type \verb;T; and
  \verb;q; proves \verb;t₁ ≃ t₂;. The point of using the term
  \verb;φ q - t₁{t₂}; at type \verb;T;, instead of the term \verb;t₁;,
  is that the \verb;φ; term erases to \verb;|t₂|;.
  Note that the types of the
  terms are not part of the equality type itself, nor does the
  elimination rule require that the types of the left-hand and right-hand
  sides are the same to do an elimination.

\item \verb;ι x : T. T';, the dependent intersection type of
  \citet{kopylov03}.  This is the type for terms \verb;t; which
  can be assigned both the type \verb;T; and the type \verb;[t/x]T';,
  the substitution instance of \verb;T'; by \verb;t;.  In the
  annotated language, we introduce a value of \verb;ι x : T. T'; by
  construct \verb;[ t, t' ];, where \verb;t; has type \verb;T;
  (algorithmically), \verb;t'; has type \verb;[t/x]T';, and
  the erasure \verb;|t|; is definitionally equal to the erasure
  \verb;|t'|;.  There are also annotated constructs
  \verb;t.1; and \verb;t.2; to select either the \verb;T; or
  \verb;[t.1/x]T'; view of a term \verb;t; of type \verb;ι x : T. T';.
\end{itemize}
It is important to understand that the described constructs are erased
before the formal reasoning (e.g. when checking if 2 terms are
definitionally equal), according to the erasure rules in
\reffig{cdle}.

\subsection{Curry-Style Typing}
\labsec{back:curry}

There is an intuitive explanation for why zero-cost (i.e. no
performance penalty) conversion should be possible
between differently indexed data (i.e. \txt{List} and \txt{Vec}) and
differently indexed programs (i.e. \txt{appL} and
\txt{appV}). In a Curry-style theory, the same
underlying untyped term can be typed multiple different
ways. Therefore, if it is possible to type a term as both a list and a
vector, then there is actually no need to do any conversion at all
because the same term can inhabit both types! In a type-annotated
(rather than type-assignment) setting, this translates to having 2
distinct terms at two distinct types, whose \textit{erasures} are equal.

\paragraph{Curry-Style Data}
As an example of Curry-style data, consider the standard definitions
of Church-encoded lists and vectors below. Note that a left-pointing
triangle (\verb;◂;) is type ascription syntax for Cedille definitions,
rather than a conventional colon (\verb;:;). The direction of the
triangle is meant to convey that the definition will be checked using
the checking (rather than inferring) mode of a bidirectional type
checker.
\begin{verbatim}
List ◂ ★ ➔ ★ = λ A. ∀ X : ★. X ➔ (A ➔ X ➔ X) ➔ X.
nilL ◂ ∀ A : ★. List A = Λ A,X. λ cN,cC. cN.
consL ◂ ∀ A : ★. A ➔ List A ➔ List A =
  Λ A. λ x,xs. Λ X. λ cN,cC. cC x (xs -X cN cC).

Vec ◂ ★ ➔ Nat ➔ ★ = λ A,n. ∀ X : Nat ➔ ★.
  X zero ➔ (∀ n : Nat. A ➔ X n ➔ X (suc n)) ➔ X n.
nilV ◂ ∀ A : ★. Vec A zero = Λ A,X. λ cN,cC. cN.
consV ◂ ∀ A : ★. ∀ n : Nat. A ➔ Vec A n ➔ Vec A (suc n) =
  Λ A,n. λ x,xs. Λ X. λ cN,cC. cC -n x (xs -X cN cC).
\end{verbatim}
Notice that the only difference between the list constructor terms
(\verb;nilL; and \verb;consL;) and vector constructor terms
(\verb;nilV; and \verb;consV;) is the number of implicit
abstractions (e.g. \verb;Λ n;) and implicit applications
(e.g. \verb;-n;). According to the erasure rules of \reffig{cdle},
this means that after erasure, \verb;nilL; and \verb;nilV; share the
same underlying untyped term
(and the same holds for \verb;consL; and \verb;consV;):
\begin{verbatim}
|nilL| = |nilV| = λ cN,cC. cN
|consL| = |consV| = λ x,xs,cN,cC. cC x (xs cN cC)
\end{verbatim}


\paragraph{Curry-Style Programs}
As an example of Curry-style programs, consider the standard
definitions of the append function for Church-encoded lists and
vectors below:
\begin{verbatim}
appL ◂ ∀ A : ★. List A ➔ List A ➔ List A
  = Λ A. λ xs. xs -(List A ➔ List A)
  (λ ys. ys)
  (Λ xs. λ x,ih,ys. consL -A x (ih ys)).
appV ◂ ∀ A : ★. ∀ n : Nat. Vec A n ➔ 
  ∀ m : Nat. Vec A m ➔ Vec A (add n m)
  = Λ A. λ xs. xs -(λ n. ∀ m : Nat. Vec A m ➔ Vec A (add n m))
  (Λ m. λ ys. ys)
  (Λ n. Λ xs,x,ih. Λ m. λ ys. consV A -(add n m) x (ih -m ys)).
\end{verbatim}
Like before, \verb;appL; and \verb;appV; only differ by implicit
abstractions and applications. An additional difference is that
\verb;appL; uses \verb;consL; in its second branch, while \verb;appV; uses
\verb;consV; in its second branch. Because (as seen above)
the erasure \verb;|consL|; is equal to the erasure \verb;|consV|;, it
follows that \verb;appL; and \verb;appV; also share the same
underlying untyped term:
\begin{verbatim}
|appL| = |appV| = 
  λ xs. xs (λ ys. ys) (λ x,ih,ys,cN,cC. cC x (ih ys cN cC))
\end{verbatim}

\subsection{Inductive Datatypes}
\labsec{back:inductive}

The enriching direction of reuse requires \textit{dependent}
function types, which must be proven by induction on their inputs using
eliminators. The Church-encoded \verb;List; and \verb;Vec; datatypes
of \refsec{back:curry} do not support induction, due to a result by
\citet{geuvers01}. However, \citet{stump18} shows that the
dependent intersection (using the $\iota$-type from \reffig{cdle})
of an impredicative Church-encoded type with a
predicate, representing what it means for the type to be inductive,
\textit{does} support induction (or an eliminator):
\begin{verbatim}
List ◂ ★ ➔ ★ = λ A. ι xs : ListChurch A. ListInductive A xs.
elimList ◂ ∀ A : ★. ∀ P : List A ➔ ★.
  P (nilL -A) ➔
  (∀ xs : List A. Π x : A. P xs ➔ P (consL -A x xs)) ➔
  Π xs : List A. P xs
\end{verbatim}
Above, \verb;ListChurch; is the renamed definition of \verb;List; from
\refsec{back:curry}. Intersection-type versions of the constructors
\verb;nilL; and \verb;consL; can also be defined. We refer readers interested
in the definitions of \verb;nilL;, \verb;consL;, and \verb;elimList;
to \citet{stump18}, as this section only depends on their
type-level interface (rather than their term-level implementation). We
also assume a corresponding $\iota$-type definition of \verb;Vec;
(in terms of \verb;VecChurch;), its constructors (\verb;nilV; and
\verb;consV;), and its eliminator (\verb;elimVec;).

An important thing to point out is that \verb;List; is
defined as the intersection of the \verb;ListChurch; and
\verb;ListInductive; types, and that intersection pairs (i.e. $[t_1,t_2]$) erase to their
first components (i.e. $t_1$ of type \verb;ListChurch;) by \reffig{cdle}.
Hence, the erased $\iota$-style \verb;nilL; is the same as the erased
Church-style \verb;nilL; (and the same holds for both styles of
\verb;consL;).

\section{The Problem \& Manual Solution}
\labsec{prob}

\refsec{back:curry} shows how differently indexed data
(e.g. \verb;List; and \verb;Vec;) and programs
(e.g. \verb;appL; and \verb;appV;) can share the same erased
untyped terms in a Curry-style dependent type theory. 
Now we consider the problem of manually reusing data and programs,
in both the forgetful (e.g. \verb;Vec; to \verb;List;, and
\verb;appV; to \verb;appL;) and enriching
(e.g. \verb;List; to \verb;Vec;, and
\verb;appL; to \verb;appV;) directions.

\subsection{The Problem: Manual Linear Time Reuse}
\labsec{prob:linear}

First, we review how to manually reuse data and programs using
linear-time conversions, which is already possible in popular dependently
typed languages. Then (in \refsec{prob:manual}),
we show how Cedille lets us manually derive
zero-cost (or constant time) conversions from the linear-time
conversions. 
After erasure, Cedille programs consist of untyped
lambda calculus terms. Hence, the cost model of Cedille is similar to
that of other untyped lambda calculus implementations, such as
Racket~\cite{lang:racket}, which is the output language of compiled
Cedille programs.
Finally, to aid readability, from now on we omit implicit
abstractions (e.g. \verb;Λ A;) and implicit applications
(e.g. \verb;-A;).\footnote{
  While we omit most implicit abstractions and applications in this
  paper, the current implementation of Cedille only supports a limited
  form of type inference. Our accompanying Cedille formalization does
  not omit any implicits.
  }

\paragraph{Linear Time Forgetful Data Reuse}
We can convert a vector to a list by iteration:
\begin{verbatim}
v2l ◂ ∀ A : ★. ∀ n : Nat. Vec A n ➔ List A
  = elimVec nilL (λ x,ih. consL x ih).
\end{verbatim}
The conversion above only requires iteration, rather than induction,
because the codomain \verb;List A; does not depend on the domain
\verb;Vec A n;. If we explicitly supplied the motive (or, predicate)
\verb;P; to \verb;elimVec;, it would ignore its argument
(i.e. \verb;P = λ xs. List A;).

\paragraph{Linear Time Enriching Data Reuse}
We can convert a list to a vector by induction:
\begin{verbatim}
l2v ◂ ∀ A : ★. Π xs : List A. Vec A (len xs)
  = elimList nilV (λ x,ih. consV x ih).
\end{verbatim}
The conversion above requires induction, rather than iteration,
because the codomain \texttt{Vec A (len xs)} depends on the domain
\verb;List A;. If we explicitly supplied the motive \verb;P; to
\verb;elimList;, it would depend on its argument
(i.e. \verb;P = λ xs. Vec A (len xs);).

\paragraph{Linear Time Forgetful Program Reuse}
After defining the type synonyms \verb;AppL; and \verb;AppV; for the
types of list and vector append, respectively, forgetful reuse of
vector append to define list append corresponds to writing a function from
\verb;AppV; to \verb;AppL;:
\begin{verbatim}
AppL ◂ ★ = ∀ A : ★. List A ➔ List A ➔ List A.
AppV ◂ ★ = ∀ A : ★. ∀ n : Nat. Vec A n ➔ 
  ∀ m : Nat. Vec A m ➔ Vec A (add n m).
appV2appL ◂ AppV ➔ AppL
  = λ appV,xs,ys. v2l (appV (l2v xs) (l2v ys)).
\end{verbatim}
The function \verb;appV2appL; first reuses vector append
(\verb;appV;) by applying \verb;appV; to the result of
translating both list arguments (\verb;xs; and \verb;ys;) to
vectors (via \verb;l2v;). Then, it translates the result of
\verb;appV; from a vector to a list (via \verb;v2l;).

\paragraph{Linear Time Enriching Program Reuse}

Enriching reuse of list append to define vector append is the
difficult direction, which requires proving a lemma stating
that once a vector has been converted to a list
(via \verb;v2l;, or forgetful data reuse), the length of the output
list is equal to (or, preserves) the length index of the input vector:
\begin{verbatim}
v2lPresLen ◂ ∀ A : ★. ∀ n : Nat. Π xs : Vec A n. n ≃ len (v2l xs) 
  = elimVec β (λ x,ih. ρ ih - β).
\end{verbatim}
Recall (from \refsec{back:cdle}) that \verb;β; is the reflexivity constructor
of an equality of type \verb;t ≃ t; for any term \verb;t;,
and that \verb;ρ; is a rewrite primitive that exchanges
occurrences of \verb;t; with occurrences of \verb;t'; in the goal,
when given evidence that \verb;t ≃ t';. The proof of \verb;v2lPresLen;
is an easy induction, which rewrites by the inductive hypothesis
(\verb;ih;) in the cons case of the input vector.

It is not possible to a reuse a function of type \verb;AppL; to define
a function of type \verb;AppV; in general, because the result of the second function
has specific index requirements (namely, that the output vector
length is the sum of the input vector lengths). Enriching
function reuse must be modulo an additional premise, which establishes
a relationship between the input and output datatype
indices.\footnote{
  Enriching data reuse of a list (\txt{List}) as a vector (\txt{Vec})
  does not require a premise, but in general enriching data reuse is
  also modulo a premise. For example, a list can only be enriched to
  an ordered list (\txt{OList}) modulo a premise that the list is
  sorted. In \refsec{rel}, we give an example of enriching a raw
  lambda-term to a typed lambda-term, provided a premise that the raw
  term is well-typed with respect to a typing relation.
  }
The premise necessary to define \verb;AppV; in terms of \verb;AppL; requires
list length (\verb;len;) to distribute through list append (\verb;appL;):
\begin{verbatim}
LenDistAppL ◂ AppL ➔ ★ = λ appL. ∀ A : ★. Π xs,ys : List A.
  add (len xs) (len ys) ≃ len (appL xs ys).
appL2appV ◂ Π appL : AppL. LenDistAppL appL ➔ AppV
  = λ appL,q,xs,ys.       // Vec A (add n m)            
  ρ v2lPresLen xs -       // Vec A (add (len (v2l xs)) m)
  ρ v2lPresLen ys -       // Vec A (add (len (v2l xs)) (len (v2l ys)))
  ρ q (v2l xs) (v2l ys) - // Vec A (len (appL (v2l xs) (v2l ys)))
  l2v (appL (v2l xs) (v2l ys)).
\end{verbatim}
After binding the arguments to \verb;appL2appV;,
the initial goal type, and the resulting goal type after each
rewrite (using \verb;ρ;), appears as a comment
(i.e. to the right of the syntactic comment delimiter \verb;//; on each line).

Initially, the length of the goal vector is the sum of the lengths of
both input vectors \verb;xs; and \verb;ys;. First, we use the
previously proven lemma \verb;v2lPresLen; to state our goal in terms
of the lists resulting from converting input vectors \verb;xs; and
\verb;ys; (via \verb;v2l;). After reusing \verb;appL; applied to both
converted list, we would like to convert the result to a vector (via
\verb;l2v;) and return it. However, the dependent data reuse function
\verb;l2v; returns a vector indexed by the \verb;len; of its input
list, but the current goal is stated in terms of a sum
(i.e. \verb;add; rather than \verb;len;). Therefore, we must first
rewrite the goal using our premise that length distributes through
append, so that we may finally return the result of applying
\verb;l2v;.

\subsection{Manual Solution: Zero-Cost Reuse}
\labsec{prob:manual}

Now we derive zero-cost (constant-time) data and program conversions
from the linear-time equivalents of \refsec{prob:linear}.
Linear time reuse
(e.g. in \refsec{prob:linear}) is already possible in conventional
Church-style type theories, but zero-cost reuse is additionally
possible in Curry-style type theories. This is semantically motivated
because a Curry-style term can inhabit multiple types,
so conversion is semantically unnecessary
(as explained in \refsec{back:curry}).
The zero-cost (data and program) conversions in this section
are all defined in two parts:
\begin{enumerate}
\item An extensional identity proof about the corresponding linear-time conversion.
\item The actual zero-cost conversion, defined using \verb;φ; from
  \reffig{cdle}, the linear-time conversion, and the extensional
  identity proof.
\end{enumerate}

\paragraph{Zero-Cost Forgetful Data Reuse}

First, we prove that the \verb;v2l; conversion is extensionally the
identity function:
\begin{verbatim}
v2lId ◂ ∀ A : ★. ∀ n : Nat. Π xs : Vec A n. v2l xs ≃ xs 
  = elimVec β (λ x,ih. ρ ih - β).
\end{verbatim}
Next, we use the \verb;φ; primitive (of \reffig{cdle}) to return the
vector input \verb;xs; at type \verb;List A;, by appealing to the
proof (\verb;v2lId;) that \verb;v2l xs; is equal to \verb;xs;.
\begin{verbatim}
v2l! ◂ ∀ A : ★. ∀ n : Nat. Vec A n ➔ List A
  = λ xs. φ (v2lId xs) - (v2l xs) {xs}.
\end{verbatim}
The \verb;φ; expression erases to the term within the braces
(\verb;{xs};) by the erasure rules of \reffig{cdle}, hence
the erasure \verb;|v2l!|; is the identity function.
Thus, \verb;v2l!; converts a vector to a list in
constant time, as applying \verb;v2l!; is definitionally equal to
applying the identity function in CDLE:
\begin{verbatim}
|v2l!| = λ xs. xs
\end{verbatim}
By convention, we suffix a
conversion function with a bang (\verb;!;) to denote its zero-cost
equivalent.

\paragraph{Zero-Cost Enriching Data Reuse}

The enriching direction of zero-cost data reuse follows the same
pattern as the forgetful direction, by first proving an extensional identity
(\verb;l2vId;), and then using it to define a zero-cost version
(\verb;l2v!;) via \verb;φ;:
\begin{verbatim}
l2vId ◂ ∀ A : ★. Π xs : List A. l2v xs ≃ xs 
  = elimVec β (λ x,ih. ρ ih - β).
l2v! ◂ ∀ A : ★. Π xs : List A. Vec A (len xs)
  = λ xs. φ (l2vId xs) - (l2v xs) {xs}.
\end{verbatim}
And similarly, \verb;l2v!; converts any list (\verb;xs;)
to a vector at zero-cost:
\begin{verbatim}
|l2v!| = (λ xs. xs)
\end{verbatim}

\paragraph{Zero-Cost Forgetful Program Reuse}

For zero-cost forgetful program reuse of vector append, we prove
the following extensional identity:
Applying the conversion \verb;appV2appL; to any implementation of
vector append (\verb;f;), \textit{and} both list arguments, is equal
to applying vector append (\verb;f;) to both list argument that have been
zero-cost converted to vectors (via \verb;l2v!;).
\begin{verbatim}
appV2appLId ◂ Π f : AppV. ∀ A : ★. Π xs,ys : List A. 
  appV2appL f xs ys ≃ f (l2v! xs) (l2v! ys)
  = λ f,xs,ys.          // v2l (f (l2v xs) (l2v ys)) ≃ f xs ys
  ρ (l2vId xs) -        // v2l (f xs (l2v ys))
  ρ (l2vId ys) -        // v2l (f xs ys)
  ρ (v2lId (f xs ys)) - // f xs ys ≃ f xs ys
  β
\end{verbatim}
The right-side of the equality in the goal begins with
\verb;f xs ys;, because the zero-cost conversions \verb;l2v! xs; and
\verb;l2v! ys; definitionally reduce to \verb;xs; and \verb;ys;,
respectively. We rewrite twice (for \verb;xs; and \verb;ys;)
by the extensional identity lemma for \verb;l2v; (using \verb;l2vId;).
Then, we rewrite once (for \verb;f xs ys;) by the extensional identity
lemma for \verb;v2l; (using \verb;v2lId;), after which our goal is
solvable by reflexivity (\verb;β;).

We define the zero-cost conversion \verb;appV2appL!; using \verb;φ; and
the identity lemma \verb;appV2appLId; applied to the vector append
argument (\verb;f;) and both list arguments (\verb;xs; and \verb;ys;):
\begin{verbatim}
appV2appL! ◂ AppV ➔ AppL
  = λ f,xs,ys. φ (appV2appLId f xs ys) - 
  (appV2appL f xs ys) {f (l2v! xs) (l2v! ys)}.
\end{verbatim}
The erased zero-cost conversion \verb;appV2appL!; also definitionally
reduces to the identity function:
\begin{verbatim}
|appV2appL!| = λ f. λ xs,ys. |f (l2v! xs) (l2v! ys)|
  = λ f. λ xs,ys. f xs ys 
  = λ f. f
\end{verbatim}
The \verb;l2v!; zero-cost conversions reduce to applications of the
identity function. Then, the body of the \verb;λ f; abstraction
$\eta$-contracts to \verb;f;, such that the entire expression reduces
to the identity function.

\paragraph{Zero-Cost Enriching Program Reuse}

The zero-cost enriching program reuse of list append requires
the following extensional identity:
Applying the conversion \verb;appL2appV; to any implementation of
list append (\verb;f;), a proof of the length distributivity premise (\verb;p;),
and both vector arguments, is equal
to applying list append (\verb;f;) to both vector argument that have been
zero-cost converted to lists (via \verb;v2l!;).
\begin{verbatim}
appL2appVId ◂ Π f : AppV. Π q : LenDistAppL f.
  ∀ A : ★. ∀ n,m : Nat. Π xs : Vec A n. Π ys : Vec A m.
  appL2appV f q xs ys ≃ f (v2l! xs) (v2l! ys)
  = λ f,q,xs,ys.        // l2v (f (v2l xs) (v2l ys)) ≃ f xs ys
  ρ (v2lId xs) -        // l2v (f xs (v2l ys)) ≃ f xs ys
  ρ (v2lId ys) -        // l2v (f xs ys) ≃ f xs ys
  ρ (l2vId (f xs ys)) - // f xs ys ≃ f xs ys
  β
\end{verbatim}
Once again, the zero-cost conversion (\verb;appL2appV!;) is defined in
terms of the linear time conversion (\verb;appL2appV;),
\verb;φ;, and the extensional identity (\verb;appL2appVId;):
\begin{verbatim}
appL2appV! ◂ Π f : AppL. LenDistAppL f ➾ AppV
  = λ f. Λ q. λ xs,ys. 
  φ (appL2appVId f q xs ys) - 
  (appL2appV f q xs ys) {f (v2l! xs) (v2l! ys)}.
\end{verbatim}
The implication (\verb;➾;) to the right of the
premise (\verb;LenDistAppL appL;) of \verb;appL2appV; is syntax for a
non-dependent implicit (or, erased) product (i.e. a \verb;∀; with no
dependency on the quantified variable).
The fact that the zero-cost conversion \verb;appL2appV!; uses an
erased premise (compared to the non-erased premise in the linear-time
conversion \verb;appL2appV;) is \textit{crucial},
allowing \verb;appL2appV!; to also erase to the identity function:
\begin{verbatim}
|appL2appV!| = (λ f. λ xs,ys. |f (v2l! xs) (v2l! ys)|) 
  = λ f. λ xs,ys. f xs ys 
  = λ f. f
\end{verbatim}
The implicit abstraction \verb;Λ q; is discarded by erasure,
allowing the erasure \verb;|appL2appV!|; to $\eta$-contract to the
identity function (similar to how \verb;|appV2appL!|; reduces).

\section{Generic Program \& Proof Reuse}
\labsec{prog}

\refsec{prob:manual} gives a zero-cost solution to the problem of
linear-time data and program reuse problem presented in
\refsec{prob:linear}. However, the reused definitions in
\refsec{prob:manual} are \textit{manually} derived. Beginning with
this section, and for the remainder of this paper, we solve the
problem of zero-cost reuse \textit{generically}.

In \refsec{prog:id} we review the type of dependent identity
functions (\verb;IdDep;), which captures a pattern appearing in the
manual zero-cost solution to reuse (\refsec{prob:manual}).
The \verb;IdDep; type is the dependent generalization of the
non-dependent \verb;Id; type introduced by \citet{firsov18b}.
\refsec{prog:fog} generically solves the
problem of \textit{forgetful} program and proof reuse, which corresponds to
defining \verb;IdDep;-closed combinators for
the type of non-dependent functions (for program reuse), and
the type of dependent functions (for proof reuse).
\refsec{prog:enr} defines 2 additional combinators to generically solve the
problem of \textit{enriching} program and proof reuse.

\subsection{Type of Dependent Identity Functions}
\labsec{prog:id}

As explained in \refsec{back:curry}, an (erased) term may have several
possible types in a Curry-style theory. Of particular importance to
our work is that the identity function, represented by the untyped
lambda term ($\lam{x}{x}$), can have many possible types. We have seen
several examples of this in \refsec{prob:linear}, where the zero-cost
conversions \verb;v2l!;, \verb;l2v!;, \verb;appV2appL!;, and
\verb;appL2appV!; all erase to the identity function. Thus, it makes
sense to define a \textit{type of dependent identity functions} for any domain
$A : \star$ and codomain $B : A \rightarrow \star$.
We informally denote the
type of dependent identity functions by $(a : A) \leq B~a$.
Inhabitance of the type $(a : A) \leq B~a$ represents
the existence of a term $\mathcal{F}$, such that
$|\mathcal{F}|=(\lam{x}{x}$), and the existence of a typing derivation
for the judgement $\Gamma\vdash \mathcal{F} : \Pi a : A.~B~a$.

\refsec{prob:linear} manually defines zero-cost conversions
using a proof that the linear-time conversion is (after erasure)
an identity operation. Hence, the zero-cost conversion depends on two parts:
\begin{enumerate}
\item The linear-time conversion.
\item A proof that the linear-time conversion is extensionally an
  identity function. 
\end{enumerate}

Now we formally derive the type of dependent identity functions $(a :
A) \leq B~a$ in Cedille as \verb;IdDep A B;,
which abstractly represents both zero-cost conversion
parts as a dependent function (\verb;Π;) returning a dependent pair
(\verb;Sigma;):\footnote{
  The dependent pair type \txt{Sigma} can be derived in Cedille just
  like the inductive \txt{List} and \txt{Vec} types, as explained in
  \refsec{back:inductive}.
  }
\begin{verbatim}
IdDep ◂ Π A : ★. Π B : A ➔ ★. ★
  = λ A,B. Π a : A. Sigma (B a) (λ b. b ≃ a).
\end{verbatim}
The type \verb;IdDep A B; is defined when \verb;A; is a type and
\verb;B; is a family of types indexed by \verb;A;. Inhabitants of
\verb;IdDep A B; take elements of (\verb;a : A;) to elements of
(\verb;b : B a;), and a proof that \verb;b; is propositionally equal
to \verb;a; (using the heterogeneous equality type
$\simeq$
from \reffig{cdle}).
We can represent the two parts more explicitly by deriving
an introduction rule that takes the (conversion) function \verb;f; and
its extensional identity proof as arguments:
\begin{verbatim}
intrIdDep ◂ ∀ A : ★. ∀ B : A ➔ ★. 
  Π f : (Π a : A. B a). (Π a : A. f a ≃ a) ➔ IdDep A B 
  =  λ f,q,a. pair (f a) (q a).
\end{verbatim}
In practice, it is more convenient to introduce elements of
\verb;IdDep; directly in terms of the underlying $\Pi\Sigma$
representation, rather than using \verb;intrIdDep;.

Now we define the crucial elimination rule \verb;elimIdDep;, which exposes
the witness $\mathcal{F}$ at type \verb;Π a : A. B a;, whose erasure
is the identity function:
\begin{verbatim}
elimIdDep ◂ ∀ A : ★. ∀ B : A ➔ ★. IdDep A B ➔ Π a : A. B a
  = λ c,a. φ (proj2 (c a)) - (proj1 (c a)) {a}.
\end{verbatim}
The elimination rule \verb;elimIdDep; uses \verb;φ; to return the input
\verb;a;, originally at type \verb;A;, at type (\verb;B a;) using the
extensional identity proof (\verb;proj2 (c a);), where \verb;c : IdDep A B;.
From the erasure rules of CDLE (in \reffig{cdle}), it follows
that for any dependent identity function \verb;c; of type \verb;IdDep A B;,
\verb;|elimIdDep c| = ; $|\mathcal{F}|$ \verb; = (λ a. a);.

Finally, notice how the definition of
\verb;elimIdDep; abstracts out a part (i.e. the use of \verb;φ; and the
extensional identity proof) of the zero-cost
conversion definitions (\verb;v2l!;, \verb;l2v!;, \verb;appV2appL!;, and
\verb;appL2appV!;) from \refsec{prob:manual}. In subsequent sections
we define \verb;IdDep;-closed combinators, taking \verb;IdDep; inputs
and producing an \verb;IdDep; output. Because the combinators always
return an \verb;IdDep;, well typed combinator definitions guarantee
the existence of zero-cost conversions
(whose witness we can always produce by applying \verb;elimIdDep;).

We will also use non-dependent identity function counterparts
\verb;Id;, \verb;intrId;, and \verb;elimId;
(of \verb;IdDep;, \verb;intrIdDep;, and \verb;elimIdDep;, respectively),
where \verb;A : ★;, but also \verb;B : ★; (rather than \verb;B : A ➔ ★;).
These are trivially derivable from the dependent versions, so we
omit their definitions. Note that our derived non-dependent \verb;Id;
type is isomorphic to the \verb;Id; type introduced by
\citet{firsov18b}. Also, note that the usage of the non-dependent
\verb;Id; type to zero-cost convert between values of different types
is similar to how the \verb;Coercible; type class~\cite{breitner+16}
is used in Haskell, and the relationship between them is further
explained in our related work (\refsec{others:hask}).

Recall our informal notation of type
\verb;IdDep A B; as $(a : A) \leq B~a$.
The informal notation is
inspired by \citet{miquel01}, who uses a non-dependent version
of this notation ($A \leq B$) for a
subtyping judgement derivable in a Curry-style theory with implicit
products. Indeed, our \verb;Id A B; is inhabited when
\verb;A; is a subtype of \verb;B;, and correspondingly all of
our combinators can also be understood as internalized subtyping
inference rules (we discuss the relationship with subtyping further in
our related work, \refsec{others:judg}).
When there is an identity function
from \verb;A; to \verb;B; (i.e. \verb;Id A B;), and both
\verb;A; and \verb;B; are functions, it becomes confusing to talk
about domains and codomains (e.g. ``domain'' could refer to the
identity function domain \verb;A;, or the
domain of the non-identity function \verb;A;,
or the domain of the non-identity function \verb;B;). To avoid confusion,
and inspired by the relationship with subtyping, we refer to the
domain of an identity function as the \textit{subtype} and the
codomain as the \textit{supertype} (thus, we can non-ambiguously refer
to the domain and codomain of the subtype, and the same for the supertype).

\subsection{Forgetful Reuse}
\labsec{prog:fog}

Now we define generic solutions to the problem of
forgetful program and proof reuse as \verb;IdDep;-closed
combinators for the non-dependent and dependent function types,
respectively. As a demonstration of using our generic solution,
we redo the \verb;appV; reuse example from \refsec{prob:manual} in
terms of our combinators (we also provide an additional example of
reusing the proof of vector append associativity).

The examples in this section assume the existence of
identity functions (i.e. values of type \verb;IdDep;)
to convert between lists and vectors:
\begin{verbatim}
v2l ◂ ∀ A : ★. ∀ n : Nat. Id (Vec A n) (List A)
l2v ◂ ∀ A : ★. IdDep (List A) (λ xs. Vec A (len xs))
\end{verbatim}
We delay the task of defining \verb;v2l; and \verb;l2v; to
\refsec{data}, where we define both identity functions as examples
of using our generic data reuse combinators.

\subsubsection{Program Reuse Combinator}

All the names of our combinators are short descriptions of their
return types. For example, below, \verb;allArr2arr; has return type
\texttt{Id (∀ i : I. X i ➔ X' i) (Y ➔ Y')}. Mnemonically,
\verb;allArr2arr; returns an identity function from \verb;allArr;
($\forall\rightarrow$) to (\verb;2;) \verb;arr; ($\rightarrow$).
We define \verb;allArr2arr; as:
\begin{verbatim}
allArr2arr ◂ ∀ I : ★. ∀ X,X' : I ➔ ★. ∀ Y,Y' : ★.
  Π r : Y ➔ I.
  Π c1 : IdDep Y (λ y. X (r y)).
  Π c2 : Π y : Y. Id (X' (r y)) Y'.
  Id (∀ i : I. X i ➔ X' i) (Y ➔ Y')
  = Λ I,X,X',Y,Y'. allPi2pi -I -X -(λ i,x. X' i) -Y -(λ y. Y').
\end{verbatim}
The combinator \verb;allArr2arr; is a generic solution to forgetful
\textit{non-dependent function} reuse (or, forgetful \textit{program}
reuse) For example, it can solve a problem like the one below, where black boxes
(\verb;●;) represent arbitrary (not necessarily the same) types:
\begin{verbatim}
Id (∀ n : Nat. Vec A n ➔ ●) (List A ➔ ●)
\end{verbatim}
The domain of the subtype is an indexed type
and the domain of the supertype is a non-indexed type.
For example, if
we were to solve the problem above with \verb;allArr2arr;, we would set
index type \verb;I; to \verb;Nat;, the type family \verb;X; to
\verb;Vec A;, and the non-indexed type \verb;Y; to \verb;List A;.
The codomains of the subtype and supertype
(i.e. the black boxes, or \verb;X'; and \verb;Y';, respectively)
cannot depend on the
explicit domain arguments (i.e. \verb;X; and \verb;Y;),
which is why we say that \verb;allArr2arr; solves the problem of
non-dependent function reuse.
However, the codomain of the subtype (\verb;X';)
can depend on the implicit index argument (of type \verb;I;).
This covers all the implicit arguments of \verb;allArr2arr;,
and now we explain the explicit arguments:

\begin{itemize}
\item The argument \verb;r;  is the
  \textit{refinement function}, computing an
  index of type \verb;I; from the
  non-indexed type \verb;Y;, e.g. \verb;len : List A ➔ Nat;.

\item The argument \verb;c1; is the
  \textit{contravariant dependent identity function}.
  It enriches the non-indexed supertype domain \verb;y : Y;,
  e.g. \verb;xs : List A;,
  to the indexed subtype domain \verb;X (r y);,
  e.g. \verb;Vec A (len xs);.
  The index is the refinement
  of the non-indexed input \verb;y;, e.g. \verb;(len xs);).

\item The argument \verb;c2; is the
  \textit{covariant non-dependent identity function}.
  It forgets the indexed subtype codomain \verb;X' (r y);
  as the non-indexed supertype codomain \verb;Y';.
\end{itemize}

Non-dependent \verb;allArr2arr; is defined in terms of its dependent
version, \verb;allPi2Pi;, which we present subsequently.

\subsubsection{Program Reuse Example}

Now we demonstrate zero-cost forgetful program reuse of vector append
to define list append, in terms of \verb;allArr2arr;. We produce
an identity function (\verb;Id;) from \verb;AppV; to \verb;AppL;,
called \verb;appV2appL; below:
\begin{verbatim}
appV2appL ◂ Id AppV AppL
  = // Id (∀ A : ★. ●) (∀ A : ★. ●)
  copyType (Λ A. 
  // Id (∀ n : Nat. Vec A n ➔ ●) (List A ➔ ●)
  allArr2arr len l2v (λ xs. 
  // Id (∀ m : Nat. Vec A m ➔ ●) (List A ➔ ●)
  allArr2arr len l2v (λ ys. 
  // Id (Vec A (add (len xs) (len ys))) (List A)
  v2l))).
\end{verbatim}
Our example includes goal types in comments, where each goal above
illustrates the part of the problem solved by a combinator
application below. The black boxes (\verb;●;) hide the parts of the
goals that are not relevant to what is being solved by the combinators
below. We begin by handling the impredicative quantification
\verb;∀ A : ★.;, which is present in both \verb;AppV; and \verb;AppL;,
using the easy to define auxiliary definition \verb;copyType; from
\reffig{aux}. Next, we apply \verb;allArr2arr; twice to handle
contravariantly enriching both arguments from lists to
vectors. In these applications, \verb;r; is the length function
(\verb;len;), and \verb;c1; is the enriching data reuse function
\verb;l2v;. Additionally, \verb;c2; becomes the remainder of the
\verb;appV2appL; definition, giving us access to the list arguments
\verb;xs; and \verb;ys;. Finally, we covariantly forget the return type from a
vector to a list using the forgetful data reuse function \verb;v2l;.

Note that \verb;appV2appL;, above, simultaneously captures the
linear-time conversion function and the extensional identity proof
from \refsec{prob:manual}
(i.e. the former \verb;appV2appL; and \verb;appV2appLId;).
We can recover the actual zero-cost
conversion by applying \verb;elimId; to our identity function:
\begin{verbatim}
appV2appL! ◂ AppV ➔ AppL = elimId appV2appL.
\end{verbatim}
Previously, we used a bang (\verb;!;) suffix as a \textit{syntactic} convention for
defining a zero-cost conversion. Now, we can also think of the elimination
rule of identity functions (\verb;elimId;) as a bang
\textit{operator}, because applying it to any \verb;Id; results in a
zero-cost conversion. From now on we omit defining the actual
zero-cost conversions (like \verb;appV2appL!;), because they can always
be recovered by applying the elimination rule for the type of identity
functions.

\subsubsection{Proof Reuse Combinator}

The combinator \verb;allPi2pi; is a generic solution to forgetful
\textit{dependent function} reuse (or, forgetful \textit{proof} reuse). For
example, it can solve a problem like the one below:
\begin{verbatim}
Id (∀ n : Nat. Π xs : Vec A n. ●) (Π xs : List A. ●)
\end{verbatim}
The subtype codomain may depend on the subtype (vector) domain, and
the supertype codomain may depend on the supertype (list) domain.
The definition of \verb;allPi2pi; follows:
\begin{verbatim}
allPi2pi ◂ ∀ I : ★. ∀ X : I ➔ ★. ∀ X' : Π i : I. X i ➔ ★. 
  ∀ Y : ★. ∀ Y' : Y ➔ ★.
  Π r : Y ➔ I.
  Π c1 : IdDep Y (λ y. X (r y)).
  Π c2 : Π y : Y. Id (X' (r y) (elimIdDep c1 y)) (Y' y).
  Id (∀ i : I. Π x : X i. X' i x) (Π y : Y. Y' y)
  = λ r,c1,c2,f. pair (λ y. elimId (c2 y) (f -(r y) (elimIdDep c1 y))) β.
\end{verbatim}
Compared to \verb;allArr2arr;, the \verb;I;, \verb;X;, \verb;Y;,
\verb;r;, and \verb;c1; arguments are the same. However, now the
subtype codomain \verb;X'; may depend on its indexed domain
\verb;X i;, and the supertype codomain \verb;Y'; may depend on its
non-indexed domain \verb;Y;. The explicit argument \verb;c2; is also
different:

\begin{itemize}
\item The argument \verb;c2; is still the \textit{covariant
  non-dependent identity function}. However, now it forgets the
  indexed subtype codomain \verb;X' (r y) (elimIdDep c1 y); as the
  non-indexed supertype codomain \verb;Y' y;. In the \verb;X';
  application, \verb;y; is zero-cost converted from \verb;Y; to
  \verb;X (r y); via the contravariant argument \verb;c1;.
\end{itemize}

Notice that \verb;allPi2pi; is reminiscent of the subtyping rule
between dependent functions. However, it is different in that it
requires the additional \verb;r; argument to handle the additional
quantification (\verb;∀ i : I;) in the subtype. Furthermore, the zero-cost
conversion \verb;elimIdDep c1 y; is explicitly performed in the type of
argument \verb;c2;, which depends on argument \verb;c1;. In a
subtyping rule the dependency and the zero-cost conversion would be
hidden, because the subsumption rule would implicitly change the type of
\verb;y; by appealing to the subtyping premise represented by \verb;c1;.

\subsubsection{Proof Reuse Example}

As an example of zero-cost proof reuse, we demonstrate how to prove
associativity of list append from the associativity of vector
append.\footnote{
  The concept of proof reuse being ``zero-cost'' may seem odd,
  as systems like Coq erase proofs during program
  extraction. However, proofs in intentional type theory may sometimes
  have computational content we wish to preserve,
  and hence it can be valuable to zero-cost reuse such proofs. For
  example, a proof that an element exists in a list can be zero-cost
  reused as a natural number index into the list.
}
First, we create type synonyms for the theorem of list append
associativity (\verb;AssocL;) and vector append associativity
(\verb;AssocV;), parameterized by a definition of list append
(\verb;AppL;) and vector append (\verb;AppV;), respectively:
\begin{verbatim}
AssocL ◂ AppL ➔ ★ = λ appL.
  ∀ A : ★. Π xs,ys,zs : List A.  
  appL (appL xs ys) zs ≃ appL xs (appL ys zs)).
AssocV ◂ AppV ➔ ★ = λ appV.
  ∀ A : ★. ∀ n : Nat. Π xs : Vec A n.
  ∀ m : Nat. Π ys : Vec A m. ∀ o : Nat. Π zs : Vec A o.
  appV (appV xs ys) zs ≃ appV xs (appV ys zs)).
\end{verbatim}
Next, we reuse any proof of \verb;AssocV; to prove
\verb;AssocL; at zero-cost:
\begin{verbatim}
assocV2assocL ◂ ∀ appV : AppV.
  Id (AssocV appV) (AssocL (elimId appV2appL appV))
  // Id (∀ A : ★. ●) (∀ A : ★. ●)
  = copyType (Λ A.
  // Id (∀ n : Nat. Π xs : Vec A n. ●) (Π xs : List A. ●)
  allPi2pi len l2v (λ xs.
  // Id (∀ m : Nat. Π ys : Vec A m. ●) (Π ys : List A. ●)
  allPi2pi len l2v (λ ys.
  // Id (∀ o : Nat. Π zs : Vec A o. ●) (Π zs : List A. ●)
  allPi2pi len l2v (λ zs.
  // Id (appV (appV xs ys) zs ≃ appV xs (appV ys zs)))
  //    (appV (appV xs ys) zs ≃ appV xs (appV ys zs)))
  id)))).
\end{verbatim}
Notice that the identity function \verb;assocV2assocL; is parameterized
by any implementation (\verb;appV;) of the type of vector append
(\verb;AppV;). We apply the type synonym for vector append associativity
(\verb;AssocV;) directly to vector append (\verb;appV;), but the type
synonym for list append associativity (\verb;AssocL;) expects an
implementation of list append (i.e. a value of type
\verb;AppL;). Hence, we apply \verb;AssocL; to the result of
zero-cost converting \verb;appV; to a list append, via
\verb;elimId appV2appL appV;, which uses our
previously defined identity function \verb;appV2appL;.

Once again, we begin solving \verb;assocV2assocL; by copying the type
parameter \verb;A; via \verb;copyType;. Next, we apply \verb;allPi2pi;
to handle the 3 primary arguments to the theorem. The final goal is
solvable by the auxiliary identity combinator for identity functions
(\verb;id; from \reffig{aux}). Before erasure, the supertype of the
final goal has instances of \verb;elimId appV2appL appV;, instead of
\verb;appV;. Similarly, before erasure, the subtype of the final goal has instances
of \verb;elimIdDep l2v xs;, instead of \verb;xs; (and similarly for
\verb;ys; and \verb;zs;).  However, because these are zero-cost
conversions, after erasure (as depicted in the comment above) the goal
is simply solvable by \verb;id;.

\subsection{Enriching Reuse}
\labsec{prog:enr}

Now we generically solve
enriching program and proof reuse as \verb;IdDep;-closed
combinators for the non-dependent and dependent function types,
respectively.
Each forgetful program and proof reuse (\refsec{prog:fog})
combinator returns a \textit{non-dependent} identity function (\verb;Id;). In
contrast, each enriching version returns a \textit{dependent}
identity function (\verb;IdDep;), where the dependency is used to
define the \textit{premise} necessary for enrichment.
We demonstrate our enriching combinators by
redoing the \verb;appL; enriching program reuse example from
\refsec{prob:manual}.

\subsubsection{Program Reuse Combinator}

The combinator \verb;arr2allArrP; is a generic solution to
\textit{enriching} non-dependent function reuse
(or, enriching \textit{program} reuse).
Recall (from \refsec{prob:manual}) that (in general)
enriching program reuse must be
performed modulo a premise required for the enrichment
to be possible.
For example, \verb;arr2allArrP; can solve a problem like the one
below:
\begin{verbatim}
IdDep (List A ➔ ●) (λ f. (Π xs : List A. ●) ➾ ∀ n : Nat. Vec A n ➔ ●)
\end{verbatim}
Program enrichment returns a dependent identity function
(\verb;IdDep;). 
An additional implicit (erased) \textit{premise} argument
(to the left of \verb;➾;)
appears in the supertype, and the premise has a dependent domain
whose type is equal to the subtype's domain (e.g. \verb;List A;).
The codomain of the premise can depend on the subtype (e.g. \verb;f;),
in addition to the domain of the premise (\verb;xs;).
The definition of \verb;arr2allArrP; follows:
\begin{verbatim}
arr2allArrP ◂ ∀ Y,Y' : ★. ∀ P : Y ➔ Y' ➔ ★. ∀ I : ★. ∀ X,X' : I ➔ ★.
  Π r : Y ➔ I.
  Π c1 : ∀ i : I. Id (X i) Y.
  Π c1' : ∀ i : I. Π x : X i. i ≃ r (elimId c1 x).
  Π c2 : ∀ i : I. Π x : X i. 
    IdDep Y' (λ y'. P (elimId c1 x) y' ➾ X' (r (elimId c1 x))).
  IdDep (Y ➔ Y') (λ f. (Π y : Y. P y (f y)) ➾ ∀ i : I. X i ➔ X' i)
  = Λ Y,Y',P,I,X,X'. pi2allPiP -Y -(λ y. Y') -P -I -X -(λ i,x. X' i).
\end{verbatim}
Enriching \verb;arr2allArrP; shares the following implicit arguments
with forgetful \verb;allArr2arr; (from \refsec{prog:fog}): \verb;Y;,
\verb;Y';, \verb;I;, \verb;X;, and \verb;X';, as well as the explicit
argument \verb;r;.  However, the premise \verb;P; appears as an
additional implicit argument of the subtype, which may depend on both
the domain (\verb;Y;) and codomain (\verb;Y';). Now we explain the
differing explicit arguments:
\begin{itemize}
\item The argument \verb;c1; is the
  \textit{contravariant non-dependent identity function}.
  It forgets the indexed supertype domain \verb;X i;,
  e.g. \verb;Vec A n;,
  as the non-indexed subtype domain \verb;Y;,
  e.g. \verb;List A;. 

\item The argument \verb;c1'; is the
  \textit{index preservation property}.
  It requires the index \verb;i; of the supertype domain
  (\verb;x : X i;) to equal the refinement (using refinement
  function \verb;r;) of zero-cost converting \verb;x;
  (using the identity function \verb;c1;),
  e.g. \verb;n ≃ len (elimId v2l xs);, where
  \verb;xs : Vec A n; and \verb;c1; is \verb;v2l;.

\item The argument \verb;c2; is the
  \textit{covariant dependent identity function}.
  It enriches the non-indexed subtype codomain \verb;Y';
  as the indexed supertype codomain \verb;X' i;.
  The enrichment codomain also gets an additional implicit premise
  argument \verb;P;, which may depend on both the subtype domain and
  codomain.
\end{itemize}

Notice that \verb;X';, in the supertype  of argument \verb;c2;, is
applied to the refinement \verb;r (elimId c1 x));,
rather than the index \verb;i;.
This makes \verb;arr2allArrP; easier to use, as the
implementation automatically rewrites by \verb;c1'; (the index
preservation property)! We point out the consequence of this automatic
rewrite in the following example.

\subsubsection{Program Reuse Example}

Below, we redo the enriching reuse of list append example from
\refsec{prob:manual}. While our forgetful function type combinators
attack 2 pieces at a time (the domains of the supertype and subtype),
the enriching function type combinators attack 3 (the additional piece
being the premise, whose type is duplicated from the subtype domain).
\begin{verbatim}
appL2appV ◂ IdDep AppL (λ appL. LenDistAppL appL ➾ AppV)
  // IdDep (∀ A : ★. ●) (λ x. (∀ A : ★. ●) ➾ ∀ A : ★. ●)
  = copyTypeP (Λ A.                          // IdDep (List A ➔ ●) 
  //   (λ f. (Π xs : List A. ●) ➾ ∀ n : Nat. Vec A n ➔ ●)
  arr2allArrP len v2l v2lPresLen (Λ n. λ xs. // IdDep (List A ➔ ●) 
  //   (λ g. (Π ys : List A. ●) ➾ ∀ m : Nat. Vec A m ➔ ●)
  arr2allArrP len v2l v2lPresLen (Λ m. λ ys.
  // IdDep (List A) (λ zs. len zs ≃ add (len xs) (len ys)
  //                       ➾ Vec A (add (len xs) (len ys)))
  subst l2v))).
\end{verbatim}
We begin with the auxiliary \texttt{copyTypeP} combinator (from
\reffig{aux}), which is a version of \texttt{copyTypeP} that also
handles the premise (i.e. the 3rd piece). Next, we use our enriching
combinator \texttt{arr2allArrP} to handle both inductive arguments of
append, to which we supply a proof of length preservation
(\verb;v2lPresLen; from \refsec{prob:linear}) as an additional
argument. This leaves us with the goal type (after erasure) in the
final comment.  Before erasure, as explained in the
\verb;assocV2assocL; example of \refsec{prog:fog},
\verb;xs; (in the supertype of the goal) is the
zero-cost conversion \verb;elimId v2l xs; (and similarly for \verb;ys;
and \verb;zs;).  Finally, we discharge the premise by rewriting, via
the auxiliary combinator \verb;subst; (from \reffig{aux}). As an
argument, \verb;subst; takes the identity function to apply after
rewriting, which is the enriching data reuse \verb;l2v; (in this
example).

The final goal type includes the sum of the lengths of both input
vectors, rather than the sum of two vector indices.
This convenience is a result
of the automatic rewriting performed in the implementation of our
\verb;arr2allArrP; combinator! In contrast, the manual definition of
\verb;appL2appV; in \refsec{prob:linear} needed to manually rewrite
by \verb;v2lPresLen; for both append inputs.

\subsubsection{Proof Reuse Combinator}

We include the definition of the enriching proof reuse combinator
(\verb;pi2allPiP;), for reference, below.  We do not describe it in
detail, as the extensions to handle the dependent arguments in
codomains \verb;X'; and \verb;Y'; follow the same pattern as
\verb;allPi2pi; from \refsec{prog:fog}.
\begin{verbatim}
pi2allPiP ◂ ∀ Y : ★. ∀ Y' : Y ➔ ★. ∀ P : Π y : Y. Y' y ➔ ★.
  ∀ I : ★. ∀ X : I ➔ ★. ∀ X' : Π i : I. X i ➔ ★.
  Π r : Y ➔ I.
  Π c1 : ∀ i : I. Id (X i) Y.
  Π c1' : ∀ i : I. Π x : X i. i ≃ r (elimId c1 x).
  Π c2 : ∀ i : I. Π x : X i. IdDep (Y' (elimId c1 x)) (λ y'.
    P (elimId c1 x) y' ➾ X' (r (elimId c1 x)) (ρ ς (c1' x) - x)).
  IdDep (Π y : Y. Y' y) (λ f. (Π y : Y. P y (f y)) ➾
    ∀ i : I. Π x : X i. X' i x)
= λ r,c1,c2,f. pair (Λ p,i. λ x. elimIdDep (ρ (c1' -i x) - c2 x)
  (f (elimId c1 x)) -(p (elimId c1 x))) β.
\end{verbatim}
Note that in the type of \verb;c2;, the second argument of the
dependent supertype codomain \verb;X'; must use \verb;c1'; to rewrite
\verb;r (elimId c1 x); to \verb;i;, because the variable \verb;x; has
type \verb;X i;. The \verb;ς; operator of Cedille takes an equality
proof of type \verb;t ≃ t';, and returns the symmetric version of type
\verb;t' ≃ t;. The implementation also exposes that the index preservation
property (\verb;c1';) is indeed automatically rewritten
(via \verb;ρ;).

We omit the example of enriching proof reuse of list append
associativity. It is very similar to the forgetful proof reuse example
of vector append associativity, because \verb;v2lPresLen; becomes an
additional argument, making the premise the trivial \verb;Unit; type.

\begin{figure}
\centering

\begin{verbatim}
id ◂ ∀ A : ★. Id A A = λ a. pair a β.
copyType ◂ ∀ F : ★ ➔ ★. ∀ G : ★ ➔ ★.
  (∀ A : ★. Id (F A) (G A)) ➔ Id (∀ A : ★. F A) (∀ A : ★. G A)
  = λ c,xs. pair (Λ A. elimId (c -A) (xs -A)) β.
copyTypeP ◂ ∀ F : ★ ➔ ★. ∀ P : Π A : ★. F A ➔ ★. ∀ G : ★ ➔ ★.
  (∀ A : ★. IdDep (F A) (λ xs. P A xs ➾ G A)) ➔
  IdDep (∀ A : ★. F A) (λ xs. (∀ A : ★. P A (xs -A)) ➾ ∀ A : ★. G A)
  = λ c,xs. pair (Λ p,A. elimIdDep (c -A) (xs -A) -(p -A)) β.
subst ◂ ∀ Y : ★. ∀ I : ★. ∀ X : I ➔ ★.∀ r : Y ➔ I. ∀ i : I.
  IdDep Y (λ y. X (r y)) ➔ IdDep Y (λ y. r y ≃ i ➾ X i)
  = λ c,y. pair (Λ q. ρ ς q - elimIdDep c y) β.
supplyPrem ◂ ∀ Y : ★. ∀ I : ★. ∀ X : I ➔ Y ➔ ★.
  IdDep Y (λ y. ∀ i : I. X i y) ➔ ∀ i : I. IdDep Y (X i)
  = λ c,y. pair (elimIdDep c y) β.
\end{verbatim}

\caption{Auxiliary identity combinator, combinators to copy a shared
  impredicative quantification, combinator to rewrite by
  an equality constraint, and combinator to supply a premise.} 
\labfig{aux}
\end{figure}

\section{Generic Data Reuse}
\labsec{data}

In this section we give the \textit{generic} zero-cost solution to the problem of
linear-time data reuse presented in \refsec{prob:linear}, and manually
solved in \refsec{prob:manual}. In \refsec{data:fix}, we review a
type of least fixed points, used to generically encode datatypes.
\refsec{data:fog} covers generic forgetful data reuse, and
\refsec{data:fog} covers generic enriching data reuse.

\subsection{Type of Least Fixed Points}
\labsec{data:fix}

\refsec{back:inductive} reviews the work by \citet{stump18} to
manually derive induction principles for Church-encoded datatypes via
intersecting (using $\iota$) with an inductivity predicate.
\citet{firsov18a} solved the same problem generically, by deriving a
least fixed point type for any \textit{functor}, composed of 4 pieces:
\begin{enumerate}
\item An object mapping (\verb;F ◂ ★ ➔ ★;).
\item An arrow mapping
  (\verb;fmap ◂ ∀ X,Y : ★. (X ➔ Y) ➔ F X ➔ F Y;).
\item A proof of the identity law for \verb;fmap;.
\item A proof of the composition law for \verb;fmap;.
\end{enumerate}

\citet{firsov18b} improved the solution by deriving a
least fixed point type that only requires 2 pieces:
\begin{enumerate}
\item A type scheme (\verb;F ◂ ★ ➔ ★;).
\item An identity mapping
  (\verb;imap ◂ ∀ X,Y : ★. Id X Y ➔ Id (F X) (F Y);).
\end{enumerate}

In type theory, the type scheme \verb;F; is the same as the object
mapping of the functor. However, the identity mapping (\verb;imap;) is
a restriction of the arrow map (\verb;fmap;), which only requires the
user to lift an identity function
(\verb;Id; from \refsec{prog:id}) between 2 types
(\verb;X; and \verb;Y;) to an identity function
between the scheme \verb;F; applied to the same 2 types.
Deriving a concrete datatype in terms of the generic encoding of
\citet{firsov18b} takes less effort (compared to using the encoding of
\citet{firsov18a}), because \verb;imap; is less onerous to define, and no
laws need to be proved.

Furthermore, the class of datatypes representable by the
\citet{firsov18b} encoding expands to include infinitary types and
positive (not merely strictly-positive) types. \citet{firsov18b} is an
``efficient'' lambda-encoding (using Mendler-style F-algebras,
described in \refsec{data:mendler}), in the
sense that inductive types support a constant-time ``predecessor''
operation (e.g. \verb;pred; for \verb;Nat;, and \verb;tail; for
\verb;List;), using only linear space in the encoding.
Expert readers may have noticed that the tail \verb;xs : List;
(where \verb;List; is Church-encoded) in
the cons case of \verb;elimList; from \refsec{back:inductive} is
erased (i.e. quantified using \verb;∀; rather than \verb;Π;), hence
computations cannot be defined with (unerased) access to the tail of the
list. Deriving induction for a concrete \verb;List; type
encoded via the work of \citet{firsov18b},
and using Mendler-style F-algebras,
solves this problem (allowing unerased quantification over the tail
via \verb;Π;, accessible in constant time).

In this work we generically solve zero-cost \emph{data reuse} by defining
combinators for the fixpoint type of \citet{firsov18b}, whose type is:
\begin{verbatim}
IdMapping ◂  (★ ➔ ★) ➔ ★ = λ F. ∀ X,Y : ★. Id X Y ➔ Id (F X) (F Y).
Fix ◂ Π F : ★ ➔ ★. IdMapping F ➔ ★
\end{verbatim}
This work derives the non-indexed fixpoint (\verb;Fix;) in terms of an
indexed fixpoint (\verb;IFix;), over indexed schemes and
index-preserving identity mappings (\verb;IIdMapping;). The
non-indexed fixpoint is the trivial case where the index is the
\verb;Unit; type (having the single inhabitant \verb;unit;). Below, we
only give the type of the indexed fixpoint \verb;IFix;, and its
implementation is a straightforward generalization of the non-indexed
version by \citet{firsov18b}:
\begin{verbatim}
IIdMapping ◂ Π I : ★. ((I ➔ ★) ➔ I ➔ ★) ➔ ★ = λ I,F. ∀ X,Y : I ➔ ★. 
  (∀ i : I. Id (X i) (Y i)) ➔ ∀ i : I. Id (F X i) (F Y i).
IFix ◂ Π I : ★. Π F : (I ➔ ★) ➔ I ➔ ★. 
  Π imap : IIdMapping I F. I ➔ ★
\end{verbatim}

\begin{figure}
\centering

\begin{verbatim}
nilLF ◂ ∀ A,X : ★. ListF A X
consLF ◂ ∀ A,X : ★. A ➔ X ➔ ListF A X
elimListF ◂ ∀ A,X : ★. ∀ P : ListF A X ➔ ★.
  P nilLF ➔
  (Π x : A. Π xs : X. P (consLF x xs)) ➔
  Π xs : ListF A X. P xs

nilVF ◂ ∀ A : ★. ∀ X : Nat ➔ ★. VecF A X zero
consVF ◂ ∀ A : ★. ∀ X : Nat ➔ ★. ∀ n : Nat. A ➔ X n ➔ VecF A X (suc n)
elimVecF ◂ ∀ A : ★. ∀ X : Nat ➔ ★. ∀ P : Π n : Nat. VecF A X n ➔ ★.
  P zero nilVF ➔
  (∀ n : Nat. Π x : A. Π xs : X n. P (suc n) (consVF x xs)) ➔ 
  ∀ n : Nat. Π xs : VecF A X n. P n xs
\end{verbatim}

\caption{Constructors and eliminators for list and vector schemes (\txt{ListF} and \txt{VecF}).}
\labfig{schemes}
\end{figure}

\subsection{Data Schemes and Identity Mappings}
\labsec{data:schemes}

The examples in the remainder of this section will demonstrate how
data reuse combinators reduce the problem of defining an identity
function between fixpoints, to defining an identity function between
schemes. This is a much simpler problem, because schemes are
essentially sums-of-products, which do \textit{not} have inductive
arguments.

Our later examples will refer to the scheme for lists (\verb;ListF;),
and the scheme for vectors (\verb;VecF;), whose Church-encodings
appear below:
\begin{verbatim}
ListF ◂ ★ ➔ ★ ➔ ★ = λ A,X. ∀ C : ★. C ➔ (A ➔ X ➔ C) ➔ C.
VecF ◂ ★ ➔ (Nat ➔ ★) ➔ Nat ➔ ★ = λ A,X,n. 
  ∀ C : Nat ➔ ★. C zero ➔ (∀ n : Nat. A ➔ X n ➔ C (suc n)) ➔ C n.
\end{verbatim}
Like \verb;Vec; from \refsec{back:curry}
(the only difference is that the inductive argument is the abstract \verb;X n; of
the scheme, rather than an inductive \verb;C n;), the cons case of
\verb;VecF; has the natural number as an implicit argument, so the
constructors of \verb;ListF; and \verb;VecF; erase to the same
underlying untyped terms. For reference, the type signatures for the
\verb;ListF; and \verb;VecF; constructors and eliminators appear in
\reffig{schemes}. We assume an intersection-type encoding of
\verb;ListF; and \verb;VecF;, using the same technique as in
\refsec{back:inductive}, to make it possible to define the
eliminators.\footnote{
  There is no predecessor problem to worry about when deriving
  induction principles (or, eliminators) for schemes, because schemes
  do not contain inductive occurrences.
}

Next, we define the identity mappings \verb;imapL; (for \verb;ListF;)
and \verb;imapV; (for \verb;VecF;), whose definitions only differ by
which eliminator is used:
\begin{verbatim}
imapL ◂ ∀ A : ★. IdMapping (ListF A) = λ c. elimListF
  (pair nilLF β) (λ x,xs. pair (consLF x (elimId c xs) β)).
imapV ◂ ∀ A : ★. IIdMapping Nat (VecF A) = λ c. elimVecF
  (pair nilVF β) (λ x,xs. pair (consVF x (elimId c xs) β)).
\end{verbatim}
The returned value is the \verb;Sigma;-type codomain of \verb;Id; from
\refsec{prog:id}, where the first component is the supertype and
second component is the equality witness. For both \verb;imapL; and
\verb;imapV;, we mostly rebuild the term with constructors. The
interesting subterm is the tail argument (\verb;elimId c xs;) of the
cons rebuilding (for both \verb;consLF; and \verb;consVF;).
In the nil cases, the second component of the pair (constructing
\verb;Sigma;) is obviously reflexivity (\verb;β;) when rebuilding
\verb;nilLF; with itself and \verb;nilVF; with itself. However, in the
cons cases, the second component is also \verb;β;. This is because the
\textit{identity} function being mapped (\verb;c;) is erased when
zero-cost converting (i.e. \verb;|elimId c xs| = xs;).
Hence, \verb;β; is evidence of the
cons rebuilding cases because \verb;|consLF x (elimId c xs)| = |consLF| x xs;,
and \verb;|consVF x (elimId c xs)| = |consVF| x xs;.

\subsection{Forgetful Reuse}
\labsec{data:fog}

\subsubsection{Data Reuse Combinator}

The combinator \verb;ifix2fix; is a generic solution to forgetful
\textit{fixpoint} reuse (or, forgetful \textit{data} reuse). For
example, it can solve a problem like the one below:
\begin{verbatim}
Id (IFix Nat (VecF A) imapV n) (Fix (ListF A) imapL)
\end{verbatim}
Above, the subtype is an indexed fixpoint and the supertype is a
non-indexed fixpoint (hence, this the forgetful direction of data
reuse). The type of \verb;ifix2fix; follows:
\begin{verbatim}
ifix2fix ◂ ∀ I : ★. ∀ F : (I ➔ ★) ➔ I ➔ ★. ∀ G : ★ ➔ ★. 
  Π imapF : IIdMapping I F. 
  Π imapG : IdMapping G.
  Π c : ∀ X : I ➔ ★. ∀ Y : ★. 
    (∀ i : I. Id (X i) Y) ➔ ∀ i : I. Id (F X i) (G Y).
  ∀ i : I. Id (IFix I F imapF i) (Fix G imapG)
\end{verbatim}
If we were to solve the problem above with \verb;ifix2fix;,
we would set index type \verb;I; to \verb;Nat;,
the indexed scheme \verb;X; to
\verb;VecF A;, and the non-indexed scheme \verb;Y; to \verb;ListF A;.
This covers all the implicit arguments of \verb;ifix2fix;,
and now we explain the explicit arguments:

\begin{itemize}
\item The argument \verb;imapF;  is the
  \textit{index-preserving identity mapping} for the indexed scheme \verb;F;,
  e.g. \verb;imapV A; for \verb;VecF A;.

\item The argument \verb;imapG; is the
  \textit{identity mapping} for the non-indexed scheme \verb;G;,
  e.g. \verb;imapL A; for \verb;ListF A;.

\item The argument \verb;c; is the \textit{identity algebra}. It
  forgets the indexed subtype scheme (\verb;F X i;) as
  the non-indexed supertype scheme (\verb;G Y;), while assuming
  how to forget the abstract indexed subtype (\verb;X;) as the
  abstract non-indexed supertype (\verb;Y;).
\end{itemize}

The type of \verb;ifix2fix; is reminiscent of standard patterns
appearing in generic programming using fixpoint encodings of
datatypes. If you define a \textit{non-recursive}
identity function between schemes, where the
``recursive'' positions \verb;X i; are abstract, and you have access to an
abstract forgetful identity function (from \verb;X i; to \verb;Y;),
you are rewarded with a \textit{recursive}
identity function between fixpoints of those schemes.

We omit the implementation of \verb;ifix2fix; combinator since the
exact details depend on a particular encoding of Mendler-style fixed
points. Intuitively, the identity function from \verb;IFix; to
\verb;Fix; is developed by using the generic dependent elimination of
\verb;IFix; to apply the \verb;c; argument on each inductive level of
the value. The premise of \verb;c;, namely
\verb;(∀ i : I. Id (X i) Y);, is the inductive hypothesis of the
dependent elimination.

\subsubsection{Data Reuse Example}

Now we demonstrate zero-cost forgetful reuse of vector data as list
data. First, we establish type synonyms for the list and vector types,
derived generically as the fixpoints of their schemes and identity
mappings:
\begin{verbatim}
List ◂ ★ ➔ ★ = λ A. Fix (ListF A) imapL.
Vec ◂ ★ ➔ Nat ➔ ★ = λ A,n. IFix Nat (VecF A) imapV n.
\end{verbatim}
Next, we define an identity function (\verb;v2l;) from
\verb;Vec A n; to \verb;List A; by applying
\verb;ifix2fix; to the identity mappings and an
identity algebra. For legibility, we provide the identity algebra
(\verb;vf2lf;) as a standalone definition:
\begin{verbatim}
vf2lf ◂ ∀ A : ★. ∀ X : Nat ➔ ★. ∀ Y :  ★. 
  Π c : ∀ n : Nat. Id (X n) Y.
  ∀ n : Nat. Id (VecF A X n) (ListF A Y)
  = λ c. elimVecF (pair nilLF β)
  (λ x,xs. pair (consLF x (elimId c xs) β)).
v2l ◂ ∀ A : ★. ∀ n : Nat. Id (Vec A n) (List A) =
  ifix2fix imapV imapL vf2lf. 
\end{verbatim}
The identity algebra \verb;vf2lf; is defined by constructing an
identity function, and the construction is very similar to how we
defined the identity mappings \verb;imapL; and \verb;imapV; in
\refsec{data:schemes}. This time, the conversion changes the types (by going
from indexed scheme \verb;VecF; to scheme \verb;ListF;), but \verb;β;
still suffices as equality in both cases because the constructors of
both schemes erase to the same untyped terms. More concretely,
\verb;|nilVF| = |nilLF|; and
\verb;|consVF x xs| = |consLF x (elimId c xs)|;. In the cons case,
\verb;xs; has (abstract vector) type \verb;X n;, but this is zero-cost
converted via \verb;c; to (abstract list) type \verb;Y;. Hence,
because we know that \verb;|consVF| = |consLF|;, it follows that:
\begin{verbatim}
|consVF x xs| = |consVF| x xs = |consLF| x xs = |consLF x (elimId c xs)|
\end{verbatim}

\subsection{Mendler-Style Algebras}
\labsec{data:mendler}

In generic developments using fixpoint-encodings of datatypes, it is
common to define non-dependent functions as the fold of an algebra.
Our generic enriching data reuse combinator (in \refsec{data:enr})
requires an algebra argument (which is folded in the dependent type signature of
the combinator). However, because our fixpoint type is defined
using a Mendler-style encoding~\cite{firsov18b}, our enriching combinator
must take a \textit{Mendler-style algebra}. Below, we give the definition of
a Mendler-style algebra (\verb;AlgM;), and we include the more familiar
Church-style algebra (\verb;AlgC;) for reference:
\begin{verbatim}
AlgC ◂ (★ ➔ ★) ➔ ★ ➔ ★ = λ F,X. F X ➔ X.
AlgM ◂ (★ ➔ ★) ➔ ★ ➔ ★ = λ F,X. ∀ R : ★. Π rec : R ➔ X. F R ➔ X.
\end{verbatim}
Mendler algebras (\verb;AlgM;) exploit parametricity to abstractly
hide inductive data via impredicative quantification
(\verb;∀ R : ★;). However, a \textit{recursion function}
(\verb;Π rec : R ➔ X;) is provided to explicitly make recursive calls
on the hidden data. \footnote{
  Mendler-style data hiding and explicit recursion is one of the
  ingredients used by \cite{firsov18b} to define constant-time
  predecessor functions.
}

Below, we give an example of defining the list length function
(\verb;len;) as the fold of a Mendler-style length algebra
(\verb;lenAlgM;). We also provide the type of the Mendler-style
\verb;foldM; function for reference.
\begin{verbatim}
lenAlgM ◂ ∀ X : ★. AlgM (ListF X) Nat
  = λ rec. elimListF zero (λ x,xs. suc (rec xs)).
len ◂ ∀ A : ★. List A ➔ Nat = foldM lenAlgM.
foldM ◂ ∀ F : ★ ➔ ★. ∀ imap : IdMapping F. ∀ X : ★. 
  AlgM F X ➔ Fix F imap ➔ X
\end{verbatim}
The length algebra (\verb;lenAlgM;) case-splits (using
\verb;elimListF;) on the scheme (\verb;ListF;) of the generically
encoded list. The nil case returns \verb;zero;, and the cons case
returns the \verb;suc;(essor) of the result of applying the recursion
function (\verb;rec;) to the abstract recursive data (\verb;xs : R;).
Our example in \refsec{data:enr} uses both the
length algebra (\verb;lenAlgM;) and the
length function (\verb;len;) defined as its fold.

\subsection{Enriching Reuse}
\labsec{data:enr}

The combinator we define in this section
(\verb;fix2ifix;) generically solves data enrichment, going from a
non-indexed to an indexed type, when the index can be
computed as a total function from the non-indexed type
(e.g. going from \verb;List; to \verb;Vec; via the total function
\verb;len : List A ➔ Nat;).

\subsubsection{Data Reuse Combinator}

Next, we define the combinator \verb;fix2ifix;, which
is a generic solution
to enriching \textit{fixpoint} reuse
(or, enriching \textit{data} reuse).
For example, it can solve a problem like the one below:
\begin{verbatim}
IdDep (Fix (ListF A) imapL)
  (λ xs. IFix Nat (VecF A) imapV (foldM lenAlgM xs))
\end{verbatim}
Notice that \verb;fix2ifix; must return a \textit{dependent} identity
function, because the index of the output vector is computed as the
length (\verb;len;) of the input list (\verb;xs;).
The type of \verb;fix2ifix; follows:
\begin{verbatim}
fix2ifix ◂ ∀ G : ★ ➔ ★. ∀ I : ★. ∀ F : (I ➔ ★) ➔ I ➔ ★.
  Π imapG : IdMapping G.
  Π imapF : IIdMapping I F. 
  Π ralg : AlgM G I. 
  Π c : ∀ Y : ★. ∀ X : I ➔ ★. Π r : Y ➔ I. 
    IdDep Y (λ y. X (r y)) ➔ 
    IdDep (G Y) (λ ys. F X (ralg -Y r ys)).
  IdDep (Fix G imapG) 
    (λ x. IFix I F imapF (foldM ralg x))
\end{verbatim}
Both \verb;fix2ifix; and \verb;ifix2fix; (from \refsec{data:fog})
share the same implicit arguments, namely \verb;I;, \verb;F;,
and \verb;G;, and they also share the explicit \verb;imapF; and
\verb;imapG; arguments. However, \verb;fix2ifix; has the following
differing explicit arguments:

\begin{itemize}
\item The argument \verb;ralg; is the
  \textit{refinement algebra} for the non-indexed scheme \verb;G;,
  e.g. \verb;lenAlgM A; for \verb;ListF A;.

\item The argument \verb;c; is the \textit{dependent identity algebra}. It
  enriches the non-indexed subtype scheme (\verb;ys : G Y;)
  to the indexed supertype scheme (\verb;F X (ralg r ys);),
  while assuming
  how to enrich the abstract non-indexed subtype (\verb;y : Y;) as
  the abstract indexed supertype (\verb;X (r y);).
\end{itemize}

Similar to the \verb;c; of forgetful \verb;ifix2fix;,
the \verb;c; of enriching \verb;fix2ifix; requires a
\textit{non-recursive} identity function between schemes, while assuming
access to an identity function between abstract ``recursive''
positions. However, the identity function in \verb;c; for
\verb;fix2ifix; are \textit{dependent}. Hence, the index of the
assumed supertype (\verb;X;) is computed from the non-indexed
subtype (\verb;y : Y;) by applying an abstract
\textit{refinement function} (\verb;r;).
Correspondingly, the index of the produced supertype
(\verb;F X;) is computed from the non-indexed
subtype (\verb;ys : G Y;) by applying the
\textit{refinement algebra} (\verb;ralg r;), while using \verb;r; for
the \verb;rec;(ursive) function of the Mendler-style algebra.

The implementation of \verb;fix2ifix;, just like \verb;ifix2fix;,
applies \verb;c; to each inductive level. The outcome is also similar,
as \verb;fix2ifix; allows the user to define a \textit{non-recursive}
identity algebra, and it produces a \textit{recursive} identity
function between fixpoints. The primary difference is
that \verb;fix2ifix; results in a dependent identity function. Hence,
the index in the dependent result is computed by folding
the Mendler-style algebra (\verb;ralg;) over the inductive input
\verb;x;. 

\subsubsection{Data Reuse Example}

Now we demonstrate zero-cost enriching reuse of list data as vector
data. The dependent identity function (\verb;l2v;)
from \verb;xs : List A; to \verb;Vec A (len xs); is defined by
applying \verb;fix2ifix; to the identity mappings and the
dependent identity algebra \verb;lf2vf;:
\begin{verbatim}
lf2vf ◂ ∀ A : ★. ∀ Y : ★. ∀ X : Nat ➔ ★.
 Π r : Y ➔ Nat. 
 Π c : IdDep Y (λ y. X (r y)).
 IdDep (ListF A Y ) (λ xs. VecF A X (lenAlgM r xs))
  = λ c. elimListF (pair nilVF β)
  (λ x,xs. pair (consVF x (elimIdDep c xs) β)).
l2v ◂ ∀ A : ★. IdDep (List A) (λ xs. Vec A (len xs))
  = fix2ifix imapL imapV lenAlgM lf2vf.
\end{verbatim}
The Mendler-style algebra used by \verb;lf2vf; is our previously
defined length algebra (\verb;lenAlgM;).
The definition of \verb;lf2vf; is essentially the same as
\verb;vf2lf; from \refsec{data:fog}, but now we eliminate a list
scheme and produce vector scheme constructors. Because the vector and
list scheme constructors erase to the same terms, the argument for why
reflexivity (\verb;β;) suffices as identity evidence stays the same.
Another difference is that the abstract tail is computed as a \textit{dependent}
elimination (\verb;elimIdDep;, rather than \verb;elimId;).
However, the dependent elimination is also erased
(\verb;|elimIdDep c xs| = xs;).

\section{Generic Relational Reuse}
\labsec{rel}

In this section we demonstrate that the techniques of our paper scale
to more complex dependently typed programs, rather than
the pedagogical list and vector running example used thus far. A now
common example (e.g. as used by \citet{tutorial:agda}) of
dependently typed programming, due to \citet{viewfromleft}, is writing
a correct-by-construction \verb;infer; function from unchecked STLC
terms (\verb;Raw;) to checked STLC terms (\verb;Term;), where
\verb;Raw; is an unindexed type and \verb;Term; is a type indexed by
the input context (\verb;Ctx;) and output type (\verb;Tp;) of
\verb;infer;. The examples of this section will use the STLC datatypes
from this problem domain:
\begin{verbatim}
Raw ◂ ★ = Fix RawF imapR.
var ◂ Nat ➔ Raw = λ n. in (varF n).
lam ◂ Tp ➔ Raw ➔ Raw = λ A,b. in (lamF A b).
app ◂ Raw ➔ Raw ➔ Raw = λ f,a. in (appF f a).

Term ◂ CtxTp ➔ ★ = IFix CtxTp TermF imapT.
ivar ◂ ∀ G : Ctx. ∀ A : Tp. Mem Tp A G ➔ Term (pair G A)
  = λ i. iin (ivarF i).
ilam ◂ ∀ G : Ctx. Π A : Tp. ∀ B : Tp. Term (pair (ext G A) B) ➔
  Term (pair G (Arr A B)) = λ A,b. iin (ilamF A b).
iapp ◂ ∀ G : Ctx. ∀ A,B : Tp. Term (pair G (Arr A B)) ➔
  Term (pair G A) ➔ Term (pair G B) = λ f,a. iin (iappF f a).
\end{verbatim}
The \verb;Raw; terms use de Bruijn indexing, hence the variable
constructor (\verb;var;) takes a natural number (\verb;Nat;). The
lambda terms are annotated so that type inference is possible, as seen
in the \verb;lam; and \verb;ilam; constructors, which use the unerased
\verb;Π; quantifer for the domain type \verb;A;. The checked
\verb;Term;'s are indexed by \verb;CtxTp;, which is simply the
non-dependent pair (\verb;Prod;, derived from \verb;Sigma;) of a
context (\verb;Ctx;, which is a \verb;List; of types \verb;Tp;) and a
type (\verb;Tp;, which can be a base type \verb;Base;, or a function
type \verb;Arr;). The indexed \verb;ivar; constructor contains a
membership proof (\verb;Mem;, a standard list membership relation),
ensuring that the type \verb;A; appears in the context
\verb;G;. Finally, the underlying schemes (\verb;RawF; and
\verb;TermF;), their identity mappings (\verb;imapR; and
\verb;imapT;), their constructors (e.g. \verb;lamF; and \verb;ilamF;),
and their eliminators (\verb;elimRawF; and \verb;elimTermF;) are
defined as in \refsec{data:schemes}, without any surprises, and have
been omitted for space reasons.

\subsection{Forgetful Data Reuse}
\labsec{rel:dfog}

Forgetful data reuse from \verb;Term; to \verb;Raw; uses
\verb;ifix2fix; (from \refsec{data:fog}) to forget
\verb;TermF; as \verb;RawF; via the helper function
\verb;tf2rf;. 
\begin{verbatim}
tf2rf ◂ ∀ X : CtxTp ➔ ★. ∀ Y : ★.
  Π c : ∀ GA : CtxTp. Id (X GA) Y.
  ∀ GA : CtxTp. Id (TermF X GA) (RawF Y)
  = λ c. elimTermF
  (λ i. pair (varF (elimId i2n i)) β)
  (λ b. pair (lamF A (elimId c b)) β)
  (λ f,a. pair (appF (elimId c f) (elimId c a)) β).
t2r ◂ ∀ GA : CtxTp. Id (Term GA) Raw
  = ifix2fix imapT imapR tf2rf.
\end{verbatim}
The lambda and application cases use the zero-cost conversion \verb;c;
to translate each \verb;X GA; to a \verb;Y;. The variable case uses the
omitted forgetful data reuse \verb;i2n; of type
\verb;Id (Mem A x xs) Nat;, which similarly forgets membership proofs
as natural numbers.

\subsection{Enriching Data Reuse}
\labsec{rel:denr}

The enriching data reuse combinator \verb;ifix2fix; (from
\refsec{data:enr}) can be used when the index of the enriched type can
be computed as a total function (via the refinement \textit{algebra}) from
the non-indexed type.
When enriching a \verb;Raw; term to a
\verb;Term;, we are faced with the problem below:
\begin{verbatim}
IdDep (Fix RawF imapR) 
  (λ t. ∀ GA : CtxTp. Typed GA t ➾ IFix CtxTp TermF imapT)
\end{verbatim}

\subsubsection{Data Reuse Combinator}

Above, \verb;Typed GA t; is an erased premise necessary for the
enrichment to be possible. To solve such a problem, we define the
generalized \verb;fix2ifixP; combinator, which allows for data
enrichment with a \textit{premise}:
\begin{verbatim}
fix2ifixP ◂ ∀ G : ★ ➔ ★. ∀ I : ★. ∀ F : (I ➔ ★) ➔ I ➔ ★.
  Π imapG : IdMapping G. Π imapF : IIdMapping I F.
  ∀ P : I ➔ FixIndM G imapG ➔ ★.
  Π c : ∀ Y : ★. ∀ X : I ➔ ★.
    Π c1 : Id Y (FixIndM G imapG).
    Π c2 : IdDep Y (λ y. ∀ i : I. P i (elimId c1 y) ➾ X i).
    IdDep (G Y) (λ ys. ∀ i : I. P i (in (elimId (imapG c1) ys)) ➾ F X i).
  IdDep (FixIndM G imapG) (λ y. ∀ i : I. P i y ➾ IFix I F imapF i)
\end{verbatim}
Besides the implicit arguments that \verb;fix2ifixP; shares with
\verb;fix2ifix;, the premise \verb;P; is an additional implicit
argument, which may dependent on the index (also implicitly quantified
as part of the premise) and the non-indexed data.

The explicit refinement algebra \verb;r; no longer appears. The
identity algebra \verb;c; generalizes to enrich the non-indexed
subtype scheme (\verb;ys : G Y;) to the indexed supertype scheme
(\verb;F X i;), once provided the index (\verb;i : I;) and the premise
(of type \verb;P i ys;) as erased arguments. These arguments must be
erased for the zero-cost conversion to indeed be an identity function
from non-indexed to indexed data. As expected in \verb;c;,
\verb;c2; is used to convert a non-indexed inductive occurrence
(\verb;Y;) to an indexed version (\verb;X i;), under an erased premise
(\verb;P i (elimId c1 y);). However, to type the premise we must
convert the abstract \verb;Y; to a concrete non-indexed
fixpoint (\verb;FixIndM G imapG;), and for this we have the
additional \verb;c1; argument.

\subsubsection{Data Reuse Example}

To enrich a \verb;Raw; term as a typed \verb;Term;, under a
well-typedness premise (\verb;Typing;), below we apply our new
combinator \verb;fix2ifixP; to the helper enriching dependent identity
algebra \verb;rf2tfP;. In many places of this paper we omit implicit
arguments, as there is a nice coincidence between when arguments are
erased and inferrable. In the example below, \verb;rf2tfP; is much
larger than previous examples because the erased index
(\verb;GA : CtxTp;) and the erased premise (of type
\verb;Typed GA (elimId c1 y);) are not inferrable. As previously, we
use dash (\verb;-;) to explicitly supply implicit arguments, but now
we use the new syntax of a variable name along with an equal sign to
only supply relevant (non-inferrable) implicit arguments
(e.g. for explicitly supplying an index and a premise,
below we use \verb;GA = ●; and \verb;e = ●;, respectively). 
\begin{verbatim}
rf2tfP ◂ ∀ Y : ★. ∀ X : CtxTp ➔ ★.
  Π c1 : Id Y Raw.
  Π c2 : IdDep Y (λ y. ∀ GA : CtxTp. Typed GA (elimId c1 y) ➾ X GA).
  IdDep (RawF Y) (λ ys. ∀ GA : CtxTp.
    Typed GA (in (elimId (imapRaw c1) ys)) ➾ TermF X GA)
  = λ c1,c2. elimRawF
  (λ n. pair (Λ GA,p. ρ (etaSigma GA) - ivarF
    (elimId n2iP n -(e = invVarLookup
      (proj1 GA) (proj2 GA) n (ρ ς (etaSigma GA) - p)))) β)
  (λ A,y. pair (Λ GC,p. ρ (etaSigma GC) - 
    ρ (invLamEq GC A (elimId c1 y) p) - ilamF
    A -(B = invLamCod GC A (elimId c1 y) p)
    (elimIdDep c2 y -(GA = pair (consL A (proj1 GC))
        (invLamCod GC A (elimId c1 y) p))
      -(e = invLamBod GC A (elimId c1 y) p))) β)
  (λ y1,y2. pair (Λ GB,p. ρ (etaSigma GB) - iappF 
    -(invAppDom GB (elimId c1 y1) (elimId c1 y2) p) -(proj2 GB)
    (elimIdDep c2 y1 -(GA = pair (proj1 GB)
        (Arr (invAppDom GB (elimId c1 y1) (elimId c1 y2) p) (proj2 GB)))
      -(e = invAppFun GB (elimId c1 y1) (elimId c1 y2) p))
    (elimIdDep c2 y2 -(GA = pair (proj1 GB)
        (invAppDom GB (elimId c1 y1) (elimId c1 y2) p))
      -(e = invAppArg GB (elimId c1 y1) (elimId c1 y2) p))) β).
r2tP ◂ IdDep Raw (λ t. ∀ GA : CtxTp. Typed GA t ➾ Term GA)
  = fix2ifixP TermF imapR imapT rf2tfP.
\end{verbatim}
Besides appealing to the $\eta$-equality of pairs (\verb;etaSigma;)
in various places for the pair of the context and the type
(e.g. \verb;GA : CtxTp;) used as the index of \verb;Term;,
\verb;rf2tfP; mostly performs case-analysis via \verb;elimRawF;, using
\verb;c2; to enrich abstract inductive \verb;Raw; terms (\verb;y : Y;)
to their typed \verb;Term; equivalents
(\verb;X GA;), once given a suitable index (\verb;GA : CtxTp;) and premise
\verb;Typed GA (elimId c1 y);.

For the lambda (\verb;ilamF;) and application (\verb;iappF;) cases,
the index (\verb;GA : CtxTp;) and \verb;Typing; premise are additional
arguments we must supply to get an identity conversion back from
\verb;c2;, which functions as an inductive hypothesis for the
purpose of this combinator. We explicitly supply both of these
implicit arguments by appealing to inversion lemmas from our
abstractly specified \verb;Typing; premise in \reffig{inv}. For
example, in the \verb;ilamF; case we get the lambda codomain via
\verb;invLamCod; and the inductive \verb;Typed; premise of the lambda
body via \verb;invLamBod;.

Note that any implementation of \verb;Typed; works for our example,
so long as the inversion lemmas are provable. Obvious choices are an
indexed datatype version of \verb;Typed; (i.e. a binary relation), or
\verb;Typed GA t; as \verb;infer (proj1 GA) t ≃ just (proj2 GA);,
making the premise state that type inference succeeds. In our
Cedille formalization, we use a third appealing ``free'' definition,
exploiting Cedille's heterogeneous equality, namely
\verb;Typed GA t; as \verb;Sigma (Term GA) (λ t'. t' ≃ t);.

Finally, the variable case (\verb;ivarF;) appeals to the omitted
enrichment \verb;n2iP; from natural numbers to membership proofs
(\verb;IdDep Nat (λ n. ∀ xs : List A. Lookup A x xs n ➾ Mem A x xs));).
The premise is an abstract \verb;Lookup; relation, of type
\verb;Π A : ★. A ➔ List A ➔ Nat ➔ ★;, stating that the element appears
at the natural number position of the (zero-indexed) list, where
\verb;Lookup; is also abstractly specified in terms of its obvious
inversion lemmas. The definition of \verb;n2iP; is similar to
\verb;r2tP;, but simpler because it only has two cases and they have
less-complicated arguments.

\begin{figure}
\centering

\begin{verbatim}
(Typed ◂ CtxTp ➔ Raw ➔ ★)
(invVarLookup ◂ Π G : Ctx. Π A : Tp. Π n : Nat. 
  Typed (pair G A) (var n) ➔ Lookup A G n)
(invLamCod ◂ Π GC : CtxTp. Π A : Tp. Π b : Raw. Typed GC (lam A b) ➔ Tp)
(invLamEq ◂ Π GC : CtxTp. Π A : Tp. Π b : Raw. Π p : Typed GC (lam A b).
  proj2 GC ≃ Arr A (invLamCod GC A b p))
(invLamBod ◂ Π GC : CtxTp. Π A : Tp. Π b : Raw. Π p : Typed GC (lam A b).
  Typed (pair (consL A (proj1 GC)) (invLamCod GC A b p)) b)
(invAppDom ◂ Π GB : CtxTp. Π f : Raw. Π a : Raw. Typed GB (app f a) ➔ Tp)
(invAppFun ◂ Π GB : CtxTp. Π f : Raw. Π a : Raw. Π p : Typed GB (app f a).
  Typed (pair (proj1 GB) (Arr (invAppDom GB f a p) (proj2 GB))) f)
(invAppArg ◂ Π GB : CtxTp. Π f : Raw. Π a : Raw. Π p : Typed GB (app f a).
  Typed (pair (proj1 GB) (invAppDom GB f a p)) a)
(termTyped ◂ Π GA : CtxTp. Π t : Term GA. Typed GA (elimId t2r t))
\end{verbatim}

\caption{An abstract typing relation,\texttt{Typed}, specified
  in terms of its inversion lemmas, and used as the premise of STLC term enrichment.}
\labfig{inv}
\end{figure}

\subsection{Enriching Program Reuse}
\labsec{rel:penr}

For our example, we show how to enrich a one-step $\beta$-reduction
function (\verb;stepR; of type \verb;StepR;) between \verb;Raw; terms to one
between typed \verb;Term;'s (of type \verb;StepT;), provided the
premise that the raw step function preserves types (\verb;TpPres;).
\begin{verbatim}
StepR ◂ ★ = Raw ➔ Raw.
TpPres ◂ StepR ➔ ★ = λ stepR. 
  Π t : Raw. Π GA : CtxTp. Typed GA t ➔ Typed GA (stepR t).
StepT ◂ ★ = ∀ GA : CtxTp. Term GA ➔ Term GA.
\end{verbatim}

\subsubsection{Data Reuse Combinator}

Recall from \refsec{prog:enr} that the enriching program reuse
combinator \verb;arr2allArrP; takes the total refinement function \verb;r;
as an argument, in addition to the index preservation property
argument \verb;c1';, which is used to automatically perform a rewrite
by the property. Below we define \verb;arr2allArrP2;, which is a
version that works with data that must be reused with a premise.
\begin{verbatim}
arr2allArrP2 ◂ ∀ Y,Y',I : ★. ∀ P : I ➔ Y ➔ ★. ∀ P' : I ➔ Y' ➔ ★.
  ∀ X,X' : I ➔ ★. Π c1 : ∀ i : I. Id (X i) Y.
  Π c1' : Π i : I. Π x : X i. P i (elimId c1 x).
  Π c2 : ∀ i : I. IdDep Y' (λ y'. P' i y' ➾ X' i).
  IdDep (Y ➔ Y') (λ f. (Π y : Y. Π i : I. P i y ➔ P' i (f y)) ➾
    ∀ i : I. X i ➔ X' i)
\end{verbatim}
Notice that the premise has been broken into two pieces (compared to
\verb;arr2allArrP;), where \verb;P; now is the premise on the
non-indexed data (of type \verb;Y;), and \verb;P'; is a premise on the
output of the function being enriched (as before). Now \verb;c1'; is a
proof that the indexed data (\verb;x : X i;) implies that the premise
holds for its forgetful variant (\verb;P i (elimId c1 x);). The
automation performed by \verb;arr2allArrP2; is automatically supplying
this evidence to the \verb;P; argument of the erased premise. Also
notice that in the erased premise, the user may assume a non-erased
(explicitly quantified) index argument (\verb;Π i : I;), because the
entire functional premise is erased anyway.

We omit the obvious dependent version (\verb;pi2allPiP2;). For space
reasons, we also omit the definition of \verb;arr2allArrP2;, which is
an obvious alteration of \verb;arr2allArrP;. In fact, in our
implementation both \verb;arr2allArrP; and \verb;arr2allArrP2; are
defined in terms of a more general version that does not perform any
automation (i.e. a combinator without a \verb;c1'; argument).

\subsubsection{Data Reuse Example}

We enrich the type-preserving one-step $\beta$-reduction function
by applying \verb;arr2allArrP2; once for the single
argument of the unary function:
\begin{verbatim}
stepR2stepT ◂ IdDep StepR (λ f. TpPres f ➾ StepT)
  // IdDep (Raw ➔ ●) (λ f. (Π t : Raw. Π GA : CtxTp. Typed GA t ➔ ●) ➾
  //   ∀ GA : CtxTp. Term GA ➔ ●)
  = arr2allArrP2 t2r termTyped
  // ∀ GA : CtxTp. IdDep Raw (λ t. Typed GA t ➾ Term GA)
  (supplyPrem r2tP).
\end{verbatim}
After applying \verb;arr2allArrP2;, the goal has the erased index part
of the argument (\verb;∀ GA : CtxTp;) on the outside of the dependent
identity function type. We solve the goal by applying the auxiliary
combinator \verb;supplyPrem; (from \reffig{aux}) to our enriching
data reuse result for terms (\verb;r2tP; from \refsec{rel:denr}),
which expects \verb;∀ GA : CtxTp; as part of the supertype premise.

Our work scales to similar but more complicated program
enrichments. See our formalization for an example of enriching a
substitution function for \verb;Raw; terms to a version for typed \verb;Term;'s,
under the premise that substitution is type-preserving. This
requires more zero-cost conversions, to go between a \verb;List; of
\verb;Raw; terms and a type-safe environment of \verb;Term;'s,
encoded as the \verb;All; type, representing that all elements of a
list satisfy some predicate.

\section{Related Work}
\labsec{others}

\subsection{Subtyping}
\labsec{others:judg}

\citet{miquel01} shows that in a Curry-style type theory with implicit
products, the subtyping \textit{judgement} can be derived as follows:
$$
\Gamma\vdash X \leq Y \triangleq \Gamma,x:X\vdash x:Y
$$

The \verb;Id; type can be seen as the internalization of this
judgement, with \verb;IdDep; corresponding to a dependent version
(i.e. our informal syntax $(x:X) \leq Y~x$, not covered by Miquel).
Miquel also showed that the subsumption rule of subtyping is
admissible in the theory with the derived judgement, and our
elimination rule \verb;elimId; corresponds to its internalization.
Finally, all of our combinators can also be translated to admissible
subtyping rules in his theory. Miquel covers several admissible
subtyping rules, but not ones corresponding to our primary forgetful
and enriching combinators for program, proof, and data reuse. Our data
reuse combinators may be of particular interest to the subtyping
community, as they corresponds to Mendler-style datatype-generic
subtyping rules.

Inspired by the internalized subtyping judgement of Miquel,
\citet{barras:implicit} show how to derive zero-cost forgetful data
reuse conversions for Church-encoded datatypes.
This work was extended by \citet{diehl} to the enriching direction.
In \refsec{prob:manual}, we derive zero-cost data reuse in terms of a
linear-time conversion and its extensional identity proof, using
$\phi$. In contrast, the zero-cost conversions of
\citet{barras:implicit} and \citet{diehl} require no extensional
identity proof, as the conversions erase to the identity function by a
clever exploitation of $\eta$-equality, without needing a rule like
$\phi$. Our work can be seen as the generic version of their manual
zero-cost reuse. When working generically with abstract combinator
definitions, an abstraction like \verb;IdDep; is necessary,
and hence also a rule like $\phi$ (used to eliminate it).



\subsection{Coercible in Haskell}
\labsec{others:hask}

Breitner et al. describe a GHC extension to Haskell (available
starting with GHC 7.8) for a type class \verb|Coercible a b|, which
allows casting from \verb|a| to \verb|b| when such a cast is indeed
the identity function~\cite{breitner+16}.  The motivation is to support retyping of data
defined using Haskell's \verb|newtype| statement, which is designed to
give programmers the power to erect abstraction barriers that cannot
be crossed outside of the module defining the \verb|newtype|.  Within
such a module, however, \verb|Coercible a b| and the associated
function \verb|coerce :: a -> b| allow programmers to apply zero-cost
casts to change between a \verb|newtype| and its definition.

\verb|Coercible| had to be added as primitive to GHC, along with a
rather complex system of \emph{roles} specifying how coercibility of
application of type constructors follows from coercibility of
arguments to those constructors.  In contrast, in the present work, we
have shown how to derive zero-cost coercions
within the existing type
theory of Cedille (via \verb;IdDep;, also derived in Cedille, which
is the dependent equivalent of \verb|Coercible|).
On the other hand, much of the
complexity of \verb|Coercible| in GHC arises from (1) how it
interoperates with programmer-specified abstraction (via
\verb|newtype|) and (2) the need to resolve \verb|Coercible a b| class
constraints automatically, similarly to other class constraints in
Haskell.  The present work does not address either issue.
However, the present work does allow for dependent casts between
indexed variants of datatypes, which \verb|Coercible| does not cover
(because \verb|Coercible| is only equivalent to our non-dependent
\verb;Id;).

\subsection{Dependent Interoperability}
\labsec{others:interop}

The field of dependent interoperability is concerned with reusing code
between non-dependent and dependent implementations of datatypes and
functions. The goal is to support interaction between non-dependent
and dependent languages, like extracted OCaml and Coq.
The most similar work to ours in this field is that of
\citet{dagand:interop}. Inspired by Homotopy Type Theory (HoTT),
\citet{dagand:interop} formalize partial equivalence types,
simultaneously representing the forgetful and enriching directions of reuse.

They also develop combinators that are closed with respect to their
partial equivalence type. For example, their \verb;HODepEquiv; combinator is
quite like our forgetful program reuse combinator
\verb;allArr2arr;. However, their work primarily focuses on the
forgetful direction of reuse for total functions, as partial functions
can be reused by inserting dynamic checks and failures using their
\textit{partial} equivalence type. In contrast, we emphasize the
\textit{total} reuse of functions in the enriching direction (like
\verb;arr2allArrP;), using premises to make the total functions
possible. Because they are primarily interested in program reuse, not
proof reuse, they do not provide dependent versions of their reuse
combinators. Additionally, they only provide combinators for function
types, not fixpoint types, as their work assumes manual solutions to
the problem of data reuse.

The class of datatypes reusable in their setting is larger, because
isomorphic datatypes, with different representations, can be
related. In contrast, our work requires the erasures of the
constructors of related types to be the same untyped terms. However,
for the price of a smaller class of reusable types, we gain the
ability to perform conversions at zero-cost.

Note that dependent interoperability~\cite{dagand:interopfoundations}
and zero-cost conversions solve
related, but different, problems. Consider the \verb;headV; function
for vectors, \texttt{headV ◂ ∀ A : ★. ∀ n : Nat. Vec A (suc n) ➔ A}.
Dependent interoperability could use this to define a partial head
function for lists, whereas we cannot perform forgetful function
when the domain is partial. At best, it would be easy to define a
forgetful function reuse combinator with a \textit{premise}, that would allow
us to reuse \verb;headV; to define
\verb;headL ◂ ∀ A : ★. ∀ n : Nat. Π xs : List A. NonEmpty xs ➾ A;
(this example would also use our data reuse combinator with a premise,
\verb;fix2ifixP;, from \refsec{rel:denr}).

Finally, \citet{dagand:interop} automates the assembly of combinators
to reuse programs by registering them as instances of Coq's type
class mechanism. Cedille does not currently have type classes, but we could
employ the same automation strategy if type classes get added to
Cedille in the future.

\subsection{Ornaments}

Ornaments~\cite{ornaments:original} are used to define refined
version of types (e.g. \texttt{Vec}) from unrefined types
(e.g. \texttt{List}) by ``ornamenting'' the unrefined type with extra
index information. In contrast, our work establishes a relationship
between \texttt{Vec} and \texttt{List} after-the-fact, by defining
forgetful and enriching \verb;IdDep; values between the types.
By \textit{defining} vectors as natural-number-ornamented lists,
ornaments can be used to calculate the ``patch'' type necessary to adapt a
function from one type to another type~\cite{ornaments:functional}.
For example, ornaments could
calculate that \texttt{LenDistAppL} is the premise necessary to adapt
\texttt{appL} from lists to vectors (\texttt{appV}).

Although ornaments can be used to derive conversions between
types in an ornamental
relationship~\cite{ornaments:original,ornaments:relational},
they take linear time, rather than constant time (i.e. the conversions
are not zero-cost).
Besides refining the indices of existing datatypes, ornaments
also allow data to be added to existing datatypes. For example, vectors can
be index-refined lists, but lists can also be natural numbers with
elements added. Our work only covers the index refinement aspect of
ornaments. It would be interesting to explore adapting a
restriction of ornaments, where only erased data can be added,
to deriving zero-cost coercions.

\subsection{Type Theory in Color}

Type Theory in Color (TTC)~\cite{bernardy:color}
generalizes the concept of erased arguments
of types to various colors, which may be erased optionally and
independently according to modalities in the type theory. In the vector
datatype declaration, the index data can be colored. If a vector is
passed to a function expecting a list (whose modality enforces the
lack of the index data color), then a forgetful zero-cost conversion
(using our parlance) is performed.

Lists can also be used as vectors, via an enriching zero-cost
conversion in the other direction. This works due to a mechanism to
interpret lists as a predicate on natural numbers. The list predicate is
generated as the erasure of its colored elements (like ornaments,
colors can add data in addition to refining indices), which results in
refining lists by the length \textit{function}.

Our work can be used to define an enriching zero-cost conversion
from natural numbers to the datatype of
finite sets (\texttt{Fin}). This is not possible with colors, because
\texttt{Fin} is indexed by successor (\texttt{suc}) in both of its
constructors, which would require generating a predicate on the
natural numbers from a non-deterministic function (or
\textit{relation}). Colors allow zero-cost conversions to be
generated and \textit{implicitly} applied because colors erase \textit{types}, as
well as values, whereas implicit products only erase values
(e.g. $\Lambda$ is erased, but not $\forall$).
Thus, while zero-cost conversions
need to be \textit{explicitly} crafted and applied in our setting, we
are able to \textit{define} zero-cost conversions (like taking natural numbers to
finite sets) for which there is no unique solution.

\section{Extensions and Future Work}
\labsec{future}

\subsection{Auxiliary Combinators}

Our program and proof reuse combinators expect index arguments to
appear next to their indexed types in type signatures. For this
reason, our combinators would not be directly applicable if we wrote
the type signature of vector append with the natural number indices of
both vector arguments at the beginning, followed by both
vectors. However, it is straightforward to define an auxiliary combinator that
flips argument order, which \citet{dagand:interop} do for their
partial type equivalence abstraction.

If a subsequent indexed type argument depends on the \textit{same}
index as a previous argument, rather than a new one, it also prevents
our combinators from being applicable. Consider the
artificial example of vector append where both input vectors must be
the same length. This can be solved via a
straightforward auxiliary combinator that introduces a new index quantification,
along with an equality that constrains the new index to equal the old index.

\subsection{Differently Indexed Combinators}

In this paper we only considered relating non-indexed types (e.g. list)
to indexed types (e.g. vector). In general, we may want to relate
an indexed type to a \textit{less indexed} type, like relating vectors
to ordered vectors in the introduction. Our combinators
straightforwardly generalize to two indexed types, \verb;X : I ➔ ★; and
\verb;Y : J ➔ ★;, along with a function that translates
the more refined index to the less refined index
(of type \verb;I ➔ J;). Our work could then be used in the common
scenario where data remains the same but only the index changes,
e.g. zero-cost converting a list of elements less than $n$, to a list
of elements less than $n + m$.

\section{Conclusion}
\labsec{conc}

We have demonstrated how to reuse programs, proofs, and types at
zero-cost, in both the forgetful and enriching directions. We achieve
this generically via combinators over the type of dependent identity
functions (\verb;IdDep;). Because partially applying the elimination
rule of \verb;IdDep; results in the term erasing to an identity
function, any conversion making use of the result of \verb;elimIdDep;
has no runtime overhead.

\begin{acks}
  We gratefully acknowledge feedback from anonymous reviewers, NSF
  support under award 1524519, and DoD support under award
  FA9550-16-1-0082 (MURI program).
\end{acks}

\bibliography{generic-reuse}



\end{document}